\documentclass[]{elsarticle} 
\usepackage{amsmath} 
\usepackage{mathtools}
\usepackage{eurosym}
\usepackage{amssymb}
\usepackage[printonlyused]{acronym}
\usepackage[labelformat = simple,format=plain,bf,font={footnotesize},margin={0.05\linewidth,0.05\linewidth},skip=8pt,tableposition=top,figureposition=bottom,singlelinecheck=false]{caption}
\usepackage[table,dvipsnames]{xcolor}
\usepackage{booktabs}
\usepackage{calc}
\usepackage[]{units}
\usepackage{array}
\usepackage{tabu}
\usepackage{longtable}
\usepackage{multirow}
\usepackage{slashbox}
\usepackage{setspace}
\usepackage[hidelinks]{hyperref}
\usepackage[hyphenbreaks]{breakurl}
\usepackage{P2defs}

\journal{Energy Economics}

\begin{document}

\begin{frontmatter}

\title{European Union gas market development}

\author[IED,IfA]{Tobias Baltensperger\corref{correspondingauthor}}
\cortext[correspondingauthor]{Corresponding author. Tel.: +41 44 632 44 73;}
\ead{t.baltensperger@usys.ethz.ch}

\author[ZHAW]{Rudolf M. F\"{u}chslin}
\author[IED]{Pius Kr\"{u}tli}
\author[IfA]{John Lygeros}

\address[IED]{Institute for Environmental Decisions (IED), ETH Z\"{u}rich, 8092 Z\"{u}rich, Switzerland}
\address[IfA]{Automatic Control Laboratory (IFA), ETH Z\"{u}rich, 8092 Z\"{u}rich, Switzerland}
\address[ZHAW]{Institute of Applied Mathematics and Physics (IAMP), ZHAW Z\"{u}rich University of Applied Sciences, 8401 Winterthur, Switzerland}

\begin{abstract}
The recently announced Energy Union by the European Commission is the most recent step in a series of developments aiming at integrating the European Union's (EU) gas markets in order to increase \sw{} and \sogs{}. Based on simulations with a spatial partial equilibrium model, we analyze the changes in consumption, prices, and \sw{} up to 2022 induced by the infrastructure expansions planned for this period. We find that wholesale prices decrease slightly and converge at Western European levels, the potential of suppliers to exert market power decreases significantly, particularly in the Baltic countries and \FI{} which are the most exposed countries today, and \cs{} increases by \unit[15.9]{\%} in the EU. 
Our results allow us to distinguish three categories of projects: (i) New gas sources developed and brought to the EU markets. These projects decrease prices and increase \sw{} in a large number of countries. The only project in this category is the Trans-Anatolian Gas Pipeline (TANAP) bringing Azeri gas to the EU; (ii) Existing gas sources made available to additional countries. This leads to an increase of \sw{} in the newly connected countries, while \sw{} drops slightly everywhere else. These projects mainly involve pipeline and regasification terminal capacity enhancements; (iii) Projects with a marginal effect on the market, assuming that it is fully functioning. Most storage expansion projects fall into this category, plus the recently announced Turkish Stream. 
Our results indicate that if all proposed infrastructure projects are realized, the EU's single market will become a reality in 2019 when \FI{} is interconnected to the EU markets. However, we also find that \sw{} can only be increased significantly for the EU as a whole if new gas sources become accessible. Consequently, we suggest that the EU should emphasize on measures to increase the available volumes, in particular once the integration of the market is completed. At the same time, efficiency gains, albeit decreasing \sw{}, help to improve the situation of consumers and decrease the dependency of the EU as a whole on external suppliers.

\end{abstract}


\end{frontmatter}

\setcounter{MaxMatrixCols}{20}
\section{Introduction} \label{sec:P2_Introduction}
The \ac{EU}\footnote{Abbreviations are listed in \mbox{Table \ref{tab:P2_Abbreviations}}.} is in a unique situation when it comes to the \sogs{}. The dependency on foreign suppliers is very high compared to other world regions with the exception of \JP{} and \KR{}; in 2011, \unit[63]{\%} of the gas consumed in the \ac{EU} was imported, compared to \unit[7]{\%} in the \US{}, \unit[22]{\%} in \CN{}, and \unit[97]{\%} in \JP{}. According to the \ac{IEA} \citep{Iea2013}, this dependency is expected to increase as a consequence of growing demand and diminishing indigenous production.
In addition, the six \ac{EU} member states \BG{}, \EE{}, \FI{}, \LT{}, \LV{}, and \SK{} import gas exclusively from \RU{}, which exposes them to supplier bargaining and disruptions \citep{EuropeanCommission2014e}. Furthermore, the markets of Eastern and Western Europe are only weakly interconnected for historical reasons. These physical limitations constrain trade in the day-to-day business and complicate, or even thwart support during a crisis.

\begin{table}[htb]
\caption{Abbreviations.}
\label{tab:P2_Abbreviations}
\centering \small \renewcommand\arraystretch{1.5}
\begin{tabu} to 0.9\linewidth {lX[l]}
\toprule
\textbf{Abbreviation} & \textbf{Description} \\
\midrule
ACER &Agency for the Cooperation of Energy Regulators\\
CEF &Connecting Europe Facility\\
EC & European Commission\\
ENTSO-G &European Network of Transmission System Operators for Gas\\
EU & European Union\\
HHI & Herfindahl-Hirschman Index\\
IEA & International Energy Agency\\
LNG &Liquefied Natural Gas\\
TANAP &Trans-Anatolian Gas Pipeline\\
TAP &Trans Adriatic Pipeline\\
TYNDP &Ten-Year Network Development Plan\\
PCI & Project of Common Interest\\
\bottomrule 
\end{tabu}
\end{table}

In 2006, Russian state-controlled Gazprom, the largest non-\ac{EU} supplier, cut off its deliveries to the \ac{EU} through Ukraine, the largest transit country (\citet{Stern2006a}), due to disputes over contractual issues with the Ukrainian state. This caused major, unprecedented disruptions in \ac{EU} gas supplies, which unmasked the vulnerability of the \ac{EU} gas markets, and sparked the development of a common approach to supply security within the \ac{EU}. The process eventually lead to the formulation of an energy strategy and the proclamation of an Energy Union by the \citet{EuropeanCommission2014e, EuropeanCommission2015}. 

The proposed energy strategy addresses the difficulties for gas supply in three ways:
(i) short-term measures, such as improving emergency mechanisms and coordination between member states in case of a disruption of supplies; 
(ii) long-term measures strengthening the internal market, such as upgrading and expanding the gas infrastructure and strengthening the regulatory framework. With the establishment of the \ac{ENTSO-G} in 2009, and the \ac{ACER} in 2011, the two key institutions of a modern gas market have been introduced to the \ac{EU} level. Furthermore, the \ac{CEF} incentivizes market forces to improve \sogs{} by identifying and co-funding otherwise unprofitable \ac{PCIs} (\citet{EuropeanCommission2014f});
and (iii) long-term measures improving the \ac{EU}'s position versus external partners. This includes promoting gas production and energy efficiency measures in the \ac{EU} to minimize overall dependency on external suppliers, diversifying external sources and developing the associated infrastructure, and supporting the contract negotiations of member states with external suppliers to make best use of the bargaining power inherent to the \ac{EU} gas market as a whole.

Although these measures were developed as a consequence of an emergency situation, they also affect the markets when undisturbed. As most of these measures will be implemented in the upcoming years, the \ac{EU} gas market is about to transform significantly. In this paper, we investigate this transformation in detail, in particular the influence of infrastructure expansions on the market under normal operation. To this end, we estimate the gas prices and consumption in the time span from 2013 to 2022 while taking projected changes in demand, production capacity, and infrastructure into account. Furthermore, we assess diversity of suppliers in each \ac{EU} member state and \CH{}\footnote{In the following we will implicitly include \CH{} when we speak about the \ac{EU} and its member states.}, the potential of suppliers to exert market power, and consumer and producer surpluses. Finally, we identify the infrastructure projects contributing most to the intended market integration and analyze them in detail. 

Our study is based on a spatial partial equilibrium model of the gas market of Europe and its main suppliers. To the best of our knowledge, this is the first study analyzing the stated aspects for Europe as a whole for the period of 2013 to 2022. We understand this work as a complement to the numerous studies emphasizing particular aspects or regions of the European gas market. Examples include \citet{Egging2008} estimating how European consumers are affected by a disruption of Russian or Algerian supplies, \citet{Lochner2012} assessing the consequences of the civil unrest in North Africa, \citet{Chyong2014} analyzing the economics of the (meanwhile canceled) South Stream project, or \citet{Cobanli2014} investigating the dynamics between the Central Asian gas producers, Russia, and the large consumer markets in Europe and China. In contrast to these studies we aim at taking a holistic view on the development of the \ac{EU} markets.

The article is structured as follows. In the next section, we introduce the model, data sources and assumptions underlying our study, and outline the calibration procedure. In \mbox{Section \ref{sec:P2_SimulationRuns}}, we describe the simulation runs conducted to assess our research questions. In \mbox{Section \ref{sec:P2_Results}}, we present our results and discuss their implications for the \ac{EU} gas market. We conclude our paper with \mbox{Section \ref{sec:P2_Conclusions}} and give an outlook on potential future research topics.

\section{Model} \label{sec:P2_ModelAndDataAssumptions}
\subsection{General setting}
The spatial partial equilibrium model introduced by \citet{Baltensperger2015} is used throughout this study. The model equations and an exemplary model with two interconnected nodes are shown in \mbox{\ref{app:P2_ModelEq_Pic_Notation}}. The model allows one to simulate the gas markets during one year, and distinguishes a summer and a winter period. We largely adopted the geographical and temporal settings on which the \ac{TYNDP} by \citet{ENTSO-G2012c} is based, since this source provides detailed data on future infrastructure projects, demand and supply affecting the \ac{EU} gas markets. We include all major producers, which supply gas to the \ac{EU}, and all \ac{EU} member states except \CY{} and \MT{} in our simulations (\mbox{Table \ref{tab:P2_ConsAndSuppl}}). Each country is modeled as a node, and the model arcs represent the total capacity of pipelines and \ac{LNG} shipments between each pair of nodes. The resulting model consists of 43 nodes and 247 arcs, and is represented by 9432 complementarity conditions. 

\begin{table}[htb]
\caption{Modeled consumers and suppliers. We group the consuming countries in ``Western European'' and ``Eastern European'' as described below. The usage of these terms is determined by the geographical location of a country and does not coincide with the historical political division of Europe. Most \ac{EU} member states produce some gas, but the volumes are generally low compared to the domestic consumption which is why we exempt them from international trade. International suppliers are marked by ${}^N$ if they are connected to the \ac{ENTSO-G} network via pipeline, and by ${}^L$ if they deliver gas to the \ac{EU} market via \ac{LNG}. }
\label{tab:P2_ConsAndSuppl} 
\centering \small \renewcommand\arraystretch{1.5}
\begin{tabu} to 0.9\linewidth {p{0.17\textwidth}X[l]}
\toprule 
Western EU consumers & \AT{}, \BE{}, \DK{}$^{N}$, \FR{}, \DE{}, \UK{}$^{N}$, \IE{}, \IT{}, \LU{}, \tNL{}$^{N}$, \PT{}, \ES{}, \SE{}, \CH{}\\
Eastern EU consumers & \BG{}, \HR{}, \CZ{}, \EE{}, \FI{}, \GR{}, \HU{}, \LV{}, \LT{}, \PL{},  \RO{}, \SK{}, \SI{} \\ 
Non-EU suppliers & \DZ{}$^{N,L}$, \AZ{}$^{N}$, \EG{}$^{L}$, \LY{}$^{N}$, Nigeria$^{L}$, \NO{}$^{N,L}$, \OM{}$^{L}$, \PE{}$^{L}$, \QA{}$^{L}$, \RU{}$^{N}$, \TT{}$^{L}$, and \YE{}$^{L}$ \\
\bottomrule 
\end{tabu}
\end{table}

In the following, we summarize the data sources used and the main assumptions made during the modeling process. We refer to \mbox{\ref{app:P2_NotationSubSec}} for an overview of the notation, and to \citet{ENTSO-G2012c} and \citet{Baltensperger2015} for additional background information.

\subsection{Modeling of consumption and prices} \label{sec:P2_DmdPrice}
The gas consumption $\sC_{nt}$ and wholesale market price $\piC_{nt}$ in each country $n$ and time period $t$ are variables of the model and are coupled by the inverse demand function $\IDF_{nt}(\sC_{nt}) = \INT_{nt} + \SLP_{nt} \cdot \sC_{nt}$, where $\piC_{nt} = \IDF_{nt}(\sC_{nt})$ in equilibrium (cf.\, Equation \eqref{eqn:P2_dpiC}). The intercept and slope are defined as:
\begin{align}
\INT_{nt}:=& (1-\frac{1}{\etaCcalib_{nt}})\cdot \piCcalib_{nt},\\
\SLP_{nt}:=&\frac{\piCcalib_{nt}}{\sCcalib_{nt} \cdot \etaCcalib_{nt}},
\end{align}
where the parameters $\sCcalib_{nt}$, $\piCcalib_{nt}$, and $\etaCcalib_{nt}$ are referred to as demand, \wtp{}, and price elasticity of demand, respectively. The parameter values are based on data of historical and estimates of future consumption $\sCdata_{nt}$, prices $\piCdata_{nt}$, price elasticities $\etaCdata_{nt}$, and the quantities $\qCdata_{fnt}$ sold by individual suppliers $f$ on the wholesale market $n$ in time period $t$. The exact values are determined in the calibration process, which is outlined in \mbox{Section \ref{sec:P2_calibration}}. In this section, we introduce the data sources of $\sCdata_{nt}$, $\piCdata_{nt}$, $\etaCdata_{nt}$, and $\qCdata_{fnt}$. 

For $\sCdata_{nt}$, we adopted the projected yearly average demands from \citet{ENTSO-G2012c}, and multiplied them by a seasonality factor, which was calculated from the monthly consumption in the years 2008-2012 reported by the \citet{EuropeanCommission} and the \citet{InternationalEnergyAgencyIEA2012c, InternationalEnergyAgencyIEA2013}.

Wholesale prices $\piCdata_{nt}$ are available for almost all \ac{EU}-countries from the \citet{EuropeanCommission2013a, EuropeanCommission2013} for the first and second quarter of 2013. We adopted the average of these prices for the year 2013 and completed the missing prices with information from the \citet{Union2014} or adopted the price of a neighboring country. For 2014-2022, only general trends are available. Hence, we used 2013-prices and adopted the average price change rate estimated for Europe in the New Policies Scenario by \citet{Iea2013}, which amounts to \unitfrac[0.23]{\%}{year}. The price development of the New Policies Scenario was chosen as a reference since the scenario's demand forecast is very similar to the one assumed in the \ac{TYNDP} \citep{ENTSO-G2012c}, on which the bulk of the remaining parameter values are based.

For the price elasticities $\etaCdata_{nt}$ we followed \citet{Lise2008}, who distinguish the three sectors industry, households and commerce, and electricity production. The sectoral price elasticities were weighted by the historical shares of the corresponding sector and summed. The shares were obtained from the \ac{TYNDP} \citep{ENTSO-G2012c} and the \citet{UnitedNations}. 

$\qCdata_{fnt}$ are only required for the calibration year, which is 2013 in our case. The data is available from the \citet{EuropeanCommission} for most \ac{EU} countries. 

Note that the wholesale markets are only modeled in \ac{EU} member states, because we concentrate on these countries in our study. For all other countries, we do not model the markets and therefore do not obtain information on consumption and prices.

\subsection{Production modeling}
Each producer outside the \ac{EU} is modeled by an individual node and produces at most the volume reported by \citet{ENTSO-G2012c} to be available for the \ac{EU}-market in 2013. This is different from the \ac{TYNDP}, where all \ac{LNG} exporters are lumped together in a single node connected to all importing countries. We generated the missing data by splitting the reported available volume of \ac{LNG} among all \ac{LNG} traders proportionally to their \ac{LNG} shares in the \ac{EU} market in 2013. The shares were obtained from \citet{BP2014}. Note that by applying this rule we exclude potential future suppliers, like the \US{}, from our calculations and instead assume that the maximum output of the current \ac{LNG} suppliers changes proportionally over the simulation horizon.

As for the \ac{EU} producers, only \tNL{}, \tUK{}, and \DK{} are modeled as international traders, while all other \ac{EU} member states are restricted to trade domestically, since these countries produce far less than they consume. We assume the production costs to be a quadratic function of the quantity produced and base our parameters on \citet{Egging2008} and \citet{InternationalEnergyAgencyIEA2012b}.

\subsection{Transportation modeling} \label{sec:P2_TransportationModeling}
The two main ways of transporting gas over long distances are by pipeline under high pressure and by ship in liquid state. Between each pair of nodes, we lump all transport capacities together, while within a node we assume that gas flow is unconstrained, partly because not all necessary data is available on a sub-country level. This assumption is sufficiently close to reality for all \ac{EU} member states except \FR{} that has a bottleneck between its northern and southern zone. This situation will not change before 2019 and might be relevant when interpreting the results.

The pipeline capacities were obtained from the \ac{TYNDP} \citep{ENTSO-G2012c} for the entire simulation horizon. We updated the assumed completion dates for the planned projects according to the \citet{EuropeanCommission2015a}. Consequently, neither Nabucco nor South Stream were included in our data set, since these projects have been canceled in the meantime. Instead, we included the recently announced Turkish Stream pipeline. The aim of the project is to install four lines with a capacity of $15.75$ billion cubic meters per year(\unitfrac[]{bcm}{y}) each. According to \citet{Stern2015}, the first pipeline is destined to supply Turkish consumers, while the others are planned to supply \ac{EU} consumers. However, due to a lack of infrastructure to transport the gas from the Greek-Turkish border to the large \ac{EU} consumer markets, the third and forth line are unlikely to be built before 2022. Hence, we only included one line (corresponding to the second line in the project) in our simulations, and assumed that line to be on-line in 2022. We will discuss the effects of this assumption in \mbox{Section \ref{sec:P2_CatOfProjects}}.

The transportation costs were derived from \citet{Chyong2014} and the information provided by the \citet{ObservatoireMediterraneendelEnergieOME2001} for pipelines outside the \ac{EU} borders. For pipelines inside the \ac{EU} borders, we adopted average tariffs charged by the national transmission system operators based on \citet{ArthurD.Little2012} and the operators' websites. Pipeline losses were set to zero since the quantities are compensated by the transmission system operator and the costs are included in the tariffs.

For the \ac{LNG} chain we made the following assumptions: For liquefaction we followed \citet[Section 3]{Shively2010} and separately accounted for fuel consumption and liquefaction cost. For \ac{LNG} shipments we assumed that the vessel is fully run with boil-off gas and assigned that quantity as a loss, which was obtained from \citet{Heede2006}. The costs were calculated as vessel rent plus a flat figure covering all other charges, including harbor fees and canal payments; these figures were obtained from \citet{TimeraEnergy2014}. Regasification costs were adopted from \citet{Chyong2014}, and losses were set to zero since they are already included in the price charged by the facility and are contained in the respective country's gas balance. 

\subsection{Storage modeling}
Storage facilities were included in the model to account for the impact of seasonality in gas consumption on the markets. The storage volumes as well as maximum injection and extraction rates were obtained from \citet{ENTSO-G2012c}. For storage costs, we followed \citet{Egging2008} and assigned costs and losses separately since the storage operator does often not compensate and charge for the losses, see for instance \citet{FluxysBelgiumSA2012}. 

\subsection{Gas traders and market power exertion} \label{sec:P2_tradersMP}
We assume that each gas producing company has an internal trading-arm which is responsible for the distribution of the gas in the domestic and foreign markets, and we call this trading-arm ``trader''. Each trader buys gas from his producer at marginal cost, and transports it to the destination markets. Pipelines outside the \ac{EU} borders are not accessible to traders other than the owner. This is particularly relevant when it comes to the Russian gas transit network, which cannot be used to transport gas between \ac{EU} countries although this would technically be feasible.

Once the gas arrives at the destination node, the trader sells it on the wholesale market. Traders can exert market power over consumers, which is modeled based on a conjectural variations approach (\citep[Chapter 12]{Tremblay2012}). The level of the market power parameter $\MP_{fnt} \in [0,1]$ determines the behavior of the traders in the market, and can be different for each trader $f$, market $n$, and time period $t$. If $\MP_{fnt}=0$, the trader $f$ is a price taker in the respective market and time period. For $\MP_{fnt}>0$, the price elasticity of the consumers in market $f$ and time period $t$ is known to trader $f$ and is taken into account when maximizing profits. $\MP_{fnt}$ is determined in the calibration process, which is outlined below.

\subsection{Calibration} \label{sec:P2_calibration} 
The model was calibrated with an earlier version of the algorithm proposed by \citet{Baltensperger2015b}, which is introduced briefly in the following. The algorithm aims at equalizing the consumption $\sC_{nt}$ to the reported values $\sCdata_{nt}$ in all nodes $n$ and time periods $t$ in 2013, and at keeping the difference between the volumes sold by each trader $f$ in each market $n$ and time period $t$ and the respective reported values low ($|\qC_{fnt} - \qCdata_{fnt}|$). These goals are linked, since the consumption is equal to the sum of sales of the traders to a particular market: $\sC_{nt} = \sum \limits_{f \in \mathcal{F}} \qC_{fnt}$.

To achieve these goals, the algorithm tunes three groups of parameters within certain bounds: (i) the \wtp{} of the consumers, $\piCcalib_{nt} \in$ [$\piCdata_{nt} \cdot 0.85$, $\piCdata_{nt} \cdot 1.15$], (ii) the price elasticity of the consumers, $\etaCcalib_{nt} \in$ [$\max\{-1, \etaCdata_{nt}-0.2\}$, $\min \{-0.3, \etaCdata_{nt}+0.2\}$], and (iii) the market power parameters of the traders, $\MP_{fnt} \in [0,1]$. As described by \citet{Baltensperger2015b}, there is a tradeoff between tight bounds on $\piCcalib_{nt}$ and $\etaCcalib_{nt}$ and low $\left|\qC_{fnt} - \qCdata_{fnt} \right| $ for all traders $f$ in a particular market $n$ and time period $t$; it is up to the modeler to balance these objectives by setting the bounds accordingly. We chose the bounds stated above to comply with the range of $\etaCcalib_{nt}$ resulting from the calibration procedure used by \citet{Chyong2014}, and to allow the typical seasonal fluctuations of $\piCcalib_{nt}$. 

Before running the calibration procedure, the reference sales of the traders $\qCdata_{fnt}$ were adjusted such that they were consistent with the consumption $\sCdata_{nt}$ and below the production capacity limit $\CAP^P_{nt}$ in each node $n$ and time period $t$. The adjusted reference sales $\qCdataApprox_{fnt}$ were deduced by proportionally reducing $\qCdata_{fnt}$ such that $\sum \limits_{f \in \mathcal{F}} \qCdataApprox_{fnt} \leq \sCdata_{nt}$ for all $n,t$, and $\sum \limits_{n^\prime \in \mathcal{N}(f)} \qCdataApprox_{f n^\prime t} \leq \overline{\CAP}^P_{n^\ast(f)t}$ for all $f \in \mathcal{F}(n),t$, where $n^\ast(f)$ is the node trader $f$ draws its gas from. In the calibration process, the $\qCdataApprox_{fnt}$ were used as lower bounds on $\qC_{fnt}$.

These adjustments cut the original values by \unit[32]{\%} on average, whereas the changes occured in countries with high re-export volumes like \BE{}. This is because re-exports are statistically recorded but are not modeled in our framework; instead, traders are in possession of the gas until they sell it on the wholesale market (\mbox{Section \ref{sec:P2_tradersMP}}). Ultimately, the low $\qCdataApprox_{fnt}$ increase the flexibility when determining feasible $\qC_{fnt}$ in the calibration process. As the flexibility is added in the \ac{EU} area in our model, which is increasingly flexible in reality anyway, we do not expect the adjustments to distort the validity of our simulation results, even though the differences between the adjusted and original values are relatively high.

\mbox{Table \ref{tab:P2_calibration}} gives an overview over the deviations of $\sCcalib_{nt}$, $\piCcalib_{nt}$, and $\etaCcalib_{nt}$, and $\qC_{fnt}$ from $\sCdata_{nt}$, $\piCdata_{nt}$, $\etaCdata_{nt}$, and $\qCdataApprox_{fnt}$. Note that seasonal data is only available for $\sCdata_{nt}$ and $\qCdataApprox_{fnt}$; for $\piCdata_{nt}$ and $\etaCdata_{nt}$ we show the yearly average values as a reference. The calibrated market power parameter values $\MP_{fnt}$ are shown in \mbox{Table \ref{tab:P2_CVvalues}}.

\begin{table}[htb]
\caption{Calibrated variables and parameters and their deviations from the reported values. For $\qC_{fnt}$, only the values for negative deviation are shown, since the available data only provides a lower limit on the flows. The figures are given in million cubic meters per day (\unitfrac{mcm}{d}) and thousand Euros per million cubic meters (\unitfrac{k\euro{}}{mcm}), respectively.}
\label{tab:P2_calibration} 
\centering \small \renewcommand\arraystretch{1.5}
\begin{tabu} to 0.9\linewidth {crrrr}
\toprule 
\textbf{Parameter/} & \multicolumn{4}{c}{\textbf{Deviation from reported value}} \\
\textbf{Variable} & \textbf{max., absolute} & \textbf{max., relative} & \textbf{mean} & \textbf{median} \\
\midrule
$\sCcalib_{nt}$ & $\unitfrac[0.99]{mcm}{d}$ & $\unit[2.38]{\%}$ & $\unitfrac[-0.02]{mcm}{d}$&$\unitfrac[0.00]{mcm}{d}$\\
$\piCcalib_{nt}$ & $\unitfrac[34.2]{k\euro{}}{mcm}$ & $\unit[13.3]{\%}$ & $\unitfrac[8.33]{k\euro{}}{mcm}$ & $\unitfrac[14.7]{k\euro{}}{mcm}$\\
$\etaCcalib_{nt}$ & $\unit[0.20]{}$ & $\unit[57.2]{\%}$ & $\unit[0.05]{}$ & $\unit[0.10]{}$\\
$\qC_{fnt}$ & $\unitfrac[17.5]{mcm}{d}$ & $\unit[100]{\%}$ & $\unitfrac[-0.21]{mcm}{d}$ & $\unitfrac[0.00]{mcm}{d}$\\
\bottomrule 
\end{tabu}
\end{table}

As \mbox{Table \ref{tab:P2_calibration}} shows, the total consumption $\sCcalib_{nt}$ and the sales per trader $\qC_{fnt}$ fit the reported data very well for all nodes $n$ and time periods $t$ in 2013, while the prices $\piCcalib_{nt}$ and price elasticities $\etaCcalib_{nt}$ remain inside their predefined bounds. 

In a final step, we derived the intercepts $\INT^C_{nt}$ and slopes $\SLP^C_{nt}$ of the affine inverse demand functions from our calibrated parameters for the markets $n$ and time periods $t$ in 2013. For 2014-2022, we assigned the consumption projected by \citet{ENTSO-G2012c} to $\sCcalib_{nt}$, adopted the growth-adjusted $\piCcalib_{nt}$, and assumed that the calibrated $\etaCcalib_{nt}$ and $\MP_{fnt}$ remain constant over the simulation horizon. 

Note that we do not model any long-term contracts, neither for \ac{LNG} nor for piped gas, as the effects on price formation are largely captured by the effects of parameter $\MP_{fnt}$. According to \citet{Smeers2008}, competition authorities argue that the overall price levels would be lower if all volumes were spot-traded, because the current lack of transparency and liquidity of the market can be used by the suppliers to ask for prices higher than marginal costs, which is exactly what we model by $\MP_{fnt}$. The only aspect of the long-term contracts which cannot be grasped with our approach is the lock-in effect over time: the model outcome changes immediately when new infrastructure goes into operation. In reality, the new infrastructure indeed puts pressure on existing contracts to adjust prices towards the new standards, but several years could pass before new contracts are put into place. In this light, the model outcome reflects the equilibrium towards which the markets converge in the long-run and cannot serve for a precise forecast of future consumption and prices.

\section{Simulation runs} \label{sec:P2_SimulationRuns}
To assess the development of the \ac{EU} gas market we carried out a series of simulations for each year over the horizon. \mbox{Table \ref{tab:P2_SimulationsRuns}} summarizes how we varied the parameters in each run, and an example for the year 2019 is given in \mbox{Table \ref{tab:P2_ImportantInfrExp2019}}. First, the status of the gas market in the previous year ($y$-1) was determined (Simulation $\SIM_{y,0}$). Then, we successively switched parameters to values of the current year ($y$) and assessed the induced changes: In $\SIM_{y,1}$, we set demand and \wtp{} in all countries to the levels of year $y$, while leaving all other parameter values at \mbox{$(y$-1)-}levels. In $\SIM_{y,2}$, we set all production and liquefaction capacities to year $y$-levels while leaving all other parameters at \mbox{$(y$-1)-}levels. In $\SIM_{y,3}$, we assessed the joint influence on the market of updated demand, \wtp{}, production and liquefaction capacities. 

\begin{table}[htb]
\caption{Parameter settings in the simulation runs. $\times$: parameter is updated to the level of year $y$. o: Parameter is not updated and is at the level of the previous year. $\mathit{C}_\mathit{all}$: Consumer behavior: assumed levels of demand and \wtp{} in all countries. $\{\mathit{P},\mathit{L}\}_\mathit{all}$: Expansion levels of production and liquefaction capacities in all countries. $\{\mathit{R},\mathit{S},\mathit{A}\}_i, i \in \{1,k_y\}$: Expansion levels of the regasification terminals, storage facilities, and pipeline capacities. $k_y$: number of infrastructure expansions in year $y$.}
\label{tab:P2_SimulationsRuns}
\centering \small \renewcommand\arraystretch{1.5}
\addtolength{\tabcolsep}{-3pt}
\begin{tabu} to 0.9\linewidth {lX[c]X[c]X[c]X[c]c}
\toprule
\textbf{Simulation}	& \multicolumn{5}{c}{\textbf{State of infrastructure}} \\
\textbf{runs} 			&$\boldsymbol{\mathit{C}_\mathit{all}}$ 	& $\boldsymbol{\{\mathit{P},\mathit{L}\}_\mathit{all}}$ & $\boldsymbol{\{\mathit{R},\mathit{S},\mathit{A}\}_1}$ & \ldots & $\boldsymbol{\{\mathit{R},\mathit{S},\mathit{A}\}_{k_y}}$\\ 
\midrule
$\SIM_{y,0}$		&o&o&o&\ldots & o	\\ 
\midrule
$\SIM_{y,1}$		&$\times$&o&o	&\ldots & o	\\ 
$\SIM_{y,2}$		&o&$\times$&o&\ldots & o		\\ 
$\SIM_{y,3}$		&$\times$&$\times$&o&\ldots & o		\\ 
\midrule
$\SIM_{y,4}$		&$\times$&$\times$&$\times$	&o & o	\\  
$\vdots$			&$\vdots$&$\vdots$&o&$\ddots$ &	o	\\ 
$\SIM_{y,k_y+3}$		&$\times$&$\times$&o&o & $\times$		\\ 
\midrule
$\SIM_{y,k_y+4}$		&$\times$&$\times$&$\times$	&\ldots & $\times$	\\ 
\bottomrule
\end{tabu}
\end{table}

For the remaining $k_y$ simulations in year $y$, the capacities of the regasification terminals, storage facilities and pipelines were updated for each country individually.
$k_y$ is the number of planned infrastructure expansions in year $y$, and varies between 8 and 27. These simulations are labeled $\SIM_{y,\ID}$, $\ID \in \{4, \ldots, k_y+3\}$. All simulations $\SIM_{y,\ID}$ also include the updated values for demand, willingness to pay, production and liquefaction capacities.
The final simulation $\SIM_{y,k_y+4}$ takes into account the effects of all changes in year $y$, and thus corresponds to $\SIM_{y+1,0}$. 

We assess the changes introduced by specific infrastructures by comparing pairs of simulation outcomes: $\SIM_{y,1}$, $\SIM_{y,2}$, and $\SIM_{y,3}$ are compared to $\SIM_{y,0}$, while all other simulations of year $y$ are compared to $\SIM_{y,3}$. Note that for $y=2013$ we only calculated $\SIM_{y,k_y+4}$, since for all other simulations the corresponding reference scenario in $y=2012$ could not be calculated due to lack of data.

We aware of the fact that this assessment is incomplete since complementary and substitution effects between infrastructure elements are not accounted for. Moreover, it is only valid for a specific point in time -- if the reference situation changes, which happens very frequently in the gas market, the assessment needs to be updated. However, these deficits could only be overcome by analyzing each infrastructure element in detail and over time, which is in conflict with our aim to capture the big picture for the European market. We accept the limitations of the chosen approach and consider them when analyzing the results.

Our study comprises 208 simulation runs in total. These were carried out on a quad-core \unit[3.4]{GHz} CPU and each took in average \unit[5]{seconds} to complete (including overhead, using MATLAB and the PATH solver), resulting in an overall simulation time of approximately \unit[17]{minutes}.

\section{Results} \label{sec:P2_Results}

\subsection{Development of consumption and market shares of suppliers} \label{sec:P2_DevOfSupplierShares}

\begin{figure*}[!htb]  
\centering
\includegraphics[width=0.9\linewidth,height=0.96\textheight,keepaspectratio]{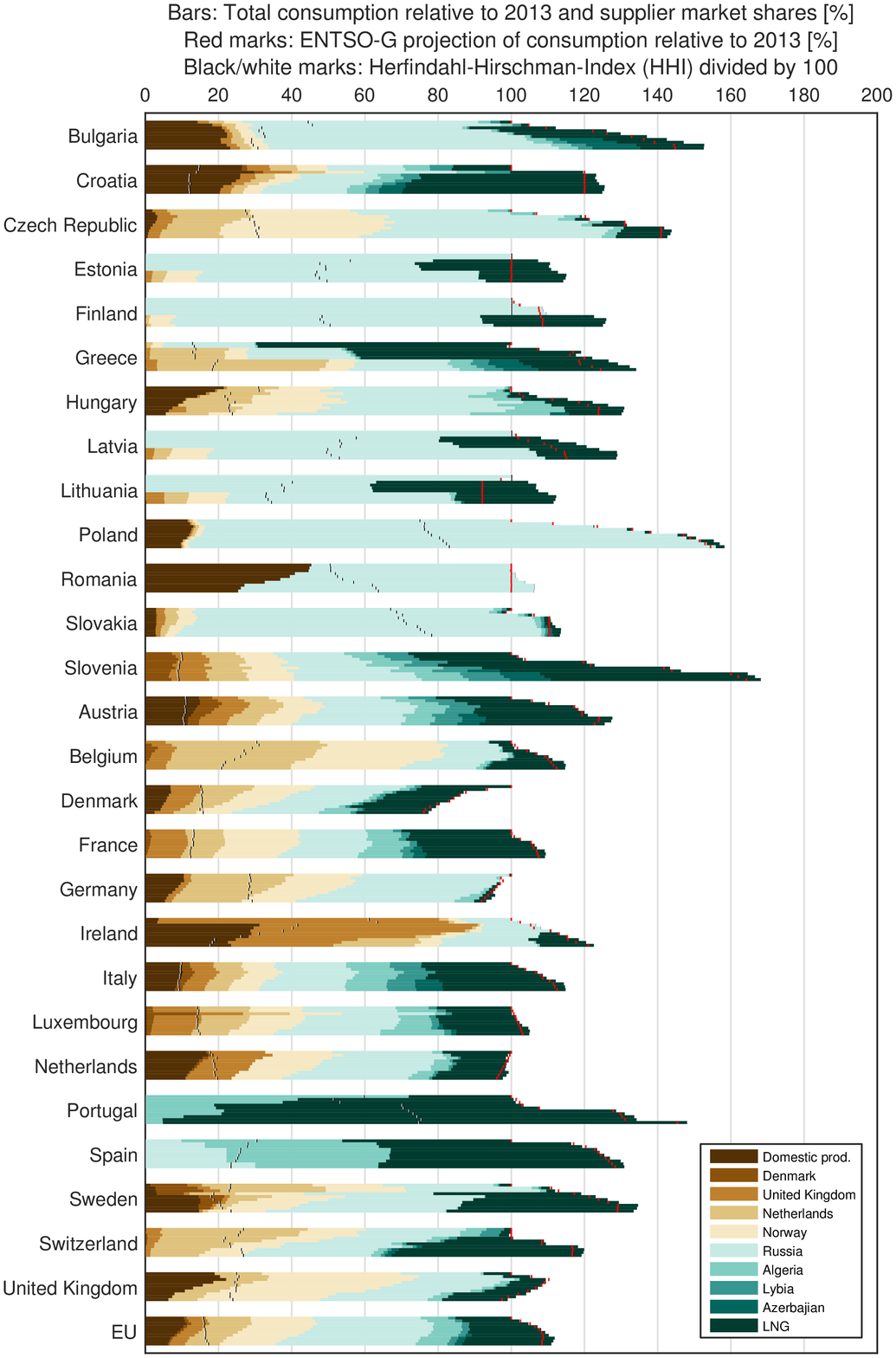}
\end{figure*}

\begin{figure*}[!htb]
\caption{Gas consumption over time and market shares of suppliers in each country and the \ac{EU} as a whole (obtained from $\SIM_{y,k_y+4}$, $y \in \{2013, \ldots, 2022\}$. For each country a group of 10 bars is displayed representing the gas consumption in the years 2013-2022 from top to bottom. The width of the bars is normalized to the consumption in 2013. Each bar is divided into at most 10 sections, with each section representing a share of a supplier in a country and year. For the \ac{EU}, ``Domestic production'' indicates the share of gas produced and consumed in the same country, while the Danish, British, and Dutch shares represent gas of those countries consumed abroad (but still within the \ac{EU}). The share of total \ac{EU} production is therefore the sum of the domestic, Danish, British, and Dutch production. The red marks indicate the demand projected by \citet{ENTSO-G2012c} over the horizon, which was taken as calibration point for the inverse demand function. The black/white marks indicate the Herfindahl-Hirschman Index (HHI) divided by 100.}
\label{fig:P2_RelConsumptionByTrader}
\end{figure*}

\mbox{Figure \ref{fig:P2_RelConsumptionByTrader}} depicts the consumption development over the simulation horizon and the market shares of the major suppliers in each \ac{EU} member state. First, we note that consumption grows in most countries and in the \ac{EU} as a whole within the next decade, with particularly high rates (\unit[$>$40]{\%}) in \PT{}, \SI{}, \CZ{}, \PL{}, and \BG{}. The growth in consumption follows the values projected by \citet{ENTSO-G2012c} for most countries; this is not surprising, since the ENTSO-G values define the inverse demand function in our model, which in turn drives the growth in consumption. Hence, interesting cases include \LT{}, for which \citet{ENTSO-G2012c} projects a decline in consumption, whereas our simulations indicate an increase, and \LV{}, \EE{}, \FI{}, \GR{}, \BG{}, and \HR{}, for which our calculations indicate a substantially larger growth than projected by \citet{ENTSO-G2012c}. These deviations originate from our more detailed representation of the effects of infrastructure development on prices and consumption: Our model shows immediate price reductions and consumption increases in regions which are newly or better connected to gas sources, whereas in the \ac{TYNDP}, consumption development for most countries is based on GDP and population growth figures, which are considerably less responsive to particular infrastructure expansions. Overall, our results indicate that \ac{EU}-wide consumption grows by \unitfrac[145]{mcm}{d} or \unit[11]{\%} over the horizon (\mbox{\ref{app:P2_ExcelTable}}), which is slightly more than projected by \citet{ENTSO-G2012c} from 2016 onwards, with a peak of $\unit[+2.9]{\%}$ (relative to the total consumption in 2013) in 2020. 

Regarding the market shares of the traders, we can observe multiple major developments. First, the share of domestic production decreases in all countries except \IE{}, \SE{}, \SK{}, and \BG{}. This is mostly driven by the growth rates in production capacities which are, similarly as the projected changes in demand, exogenous to the model and are negative for most \ac{EU} countries. This induces the second major trend: imports from non-\ac{EU} countries increase significantly, particularly from \RU{} and \ac{LNG} traders, and after 2019 from \AZ{}, to make up for the increase in consumption and the decline in domestic production. The third major trend is the diversification of suppliers in some highly exposed countries: the Baltic states, \FI{}, and \BG{}. Starting in 2015, the Baltic states and \BG{} reduce their dependency on Russian gas by increasing imports from other suppliers; \FI{} follows the same path starting in 2019. To quantify this trend, we calculate the \ac{HHI}, which is defined as the sum of the squares of the market shares of the suppliers in a market, and is measure of market concentration. Although the \ac{HHI} falls below $6000$ in the Baltic states and \FI{}, and to approximately $3000$ in \BG{}, these markets remain highly concentrated\footnote{The \citet{DoJ2009} characterize markets with a \ac{HHI} below $1500$ as unconcentrated, between $1500$ and $2500$ as moderately concentrated, and above $2500$ as highly concentrated.}. Other Eastern European countries, such as \RO{}, \SK{}, and \PL{}, experience a contraction of supplier diversity, mainly because the growing gap between consumption and production is covered by additional Russian imports, which were already high before, and are among the countries with the highest market concentrations by 2022. Lastly, note that the \ac{HHI} of the \ac{EU} as a whole remains roughly constant at a moderate level implying that diversification of suppliers is a regional problem in the first place.

\subsection{Price composition development and market power exertion} \label{sec:P2_PriceDev}

\begin{figure*}[!htb]  
\centering
\includegraphics[width=0.9\linewidth,height=0.96\textheight,keepaspectratio]{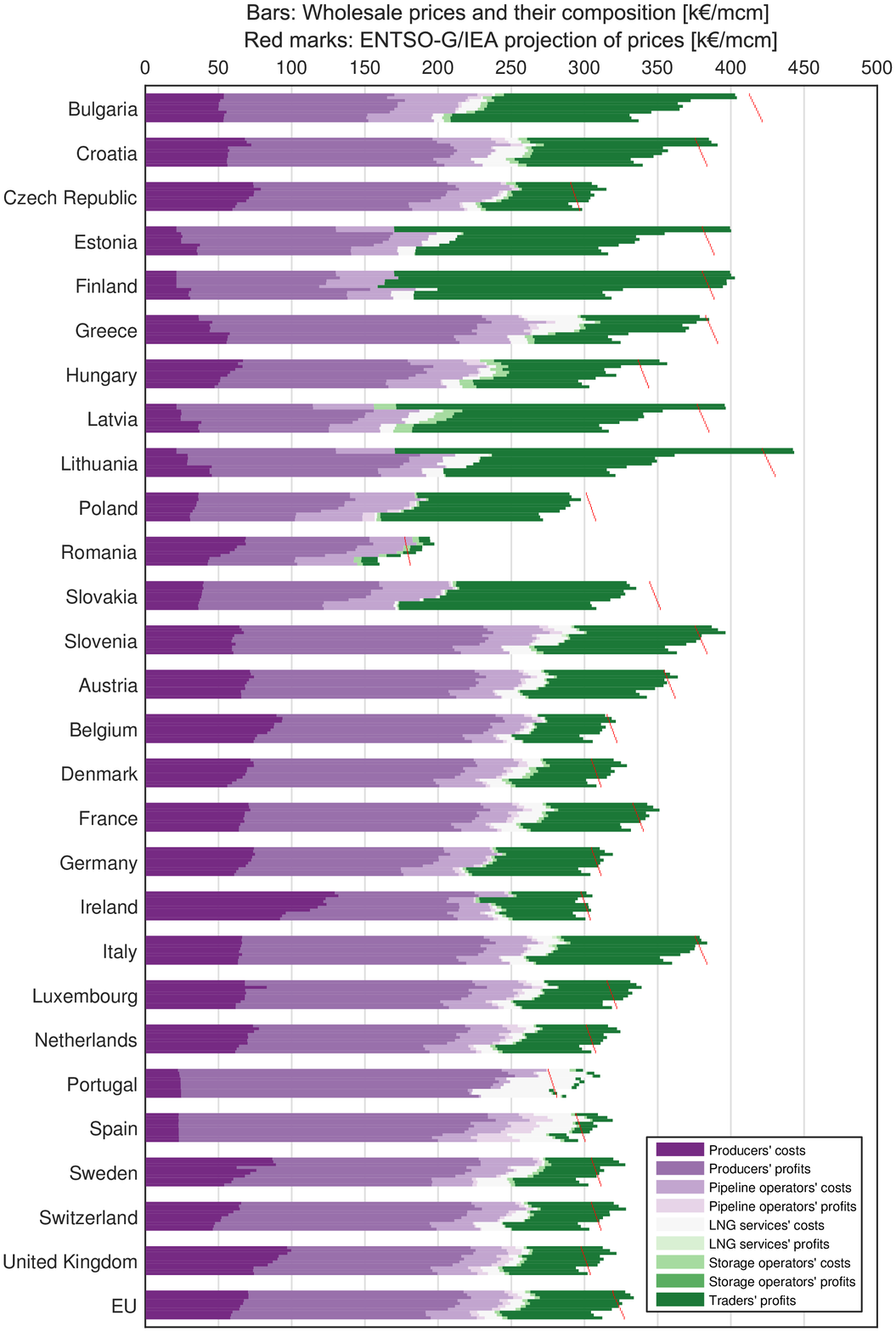}
\end{figure*}

\begin{figure*}[!htb]  
\caption{Wholesale price composition over time in the \ac{EU}. For each country, a series of 10 bars is displayed representing the prices in 2013-2022 from top to bottom. Each bar is divided into at most 9 sections. Each section represents a fraction of the costs and indicates at which stage in the supply chain it arises. The red marks indicate the price forecasts as introduced in \mbox{Section \ref{sec:P2_DmdPrice}}.
}
\label{fig:P2_PriceDev} 
\end{figure*}
\mbox{Figure \ref{fig:P2_PriceDev}} displays the price development in 2013-2022. In most markets, the prices follow a similar pattern: they first increase and peak in 2015, then decrease until 2020, and finally increase again slightly. These movements reflect the ratio of total available production capacity to total demand over time. Eastern European countries experience significant infrastructure expansions in transportation volumes relative to their consumed quantities. As a consequence, prices drop more significantly in Eastern than in Western Europe, and the prices converge to a level of \unitfrac[300-320]{k\euro{}}{mcm} by 2022. This overall trend contrasts the generally slightly increasing prices predicted by the \citet{Iea2013}, but is, however, in line with our previous results which indicated a larger than predicted consumption (\mbox{Section \ref{sec:P2_DevOfSupplierShares}}).

Wholesale prices comprise three types of costs: (i) the costs arising throughout the supply chain from production and service usage; (ii) the profits of the producers and service providers, originating from the fact that the price levels are above marginal costs for most of the gas produced; and (iii) the profits of the traders, which are greater than zero if traders exert market power over consumers, and zero otherwise. 
The price components can be calculated for every given situation as they are a function of the market equilibrium and the associated costs of each stage in the supply chain. However, as soon as the sales of the traders in the individual markets change, the cost calculation has to be redone. As we see from \mbox{Figure \ref{fig:P2_PriceDev}}, the movements are smooth over the time horizon, however, they are largely unpredictable for an alternative situation without rerunning the model. This implies that the results cannot be extrapolated to similar market situations; for instance, if none of the traders would exert market power (all else equal), prices would not fall exactly by the traders' profits. Prices would indeed fall, because traders have no incentive to withhold gas from consumers and therefore the overall available gas increases. However, we would also see two additional effects: On one hand, lower prices increase consumption, and thus production and network congestion; On the other hand, perfectly competitive traders have no incentive to diversify consumers (\citet{Baltensperger2015}) and therefore are likely to ship gas to nearby markets to save transport costs, which reduces network congestion. How the combination of these effects influences the wholesale price can only be quantified by rerunning the model. 

As \mbox{Figure \ref{fig:P2_PriceDev}} shows, the largest price components are the producers' profits, the traders' profits, and in some regions the producers' costs. The other services' costs are comparatively low, and their profits negligible. Furthermore, the price composition varies from region to region, and changes significantly over time. In the Baltic countries and \FI{} the traders' profits are the highest throughout the \ac{EU} and make up for more than \unit{50}{\%} of the final price in the first years of the simulation horizon. In 2015 and 2019, respectively, these countries are connected to the global gas market after a regasification terminal is commissioned in \LT{} in 2015, and a connection to the \ac{ENTSO-G} network is established in 2019 (see \mbox{\ref{app:P2_ExcelTable}} for all the infrastructure changes). This allows traders other than \RU{} to enter the market, as illustrated in \mbox{Figure \ref{fig:P2_RelConsumptionByTrader}}. 

At the other extreme, the largest part of the prices on the Iberian peninsula are made up from the producers profits. This is a consequence of the market clearing at the marginal costs of the most expensive producer (adjusted by transportation and storage costs), and production costs varying significantly among the producers. 

For most other countries, the costs are more evenly distributed along the supply chain. The fractions of the producers' costs and profits are larger in Western than Eastern Europe, because most local gas supplies are more expensive to produce and distances to low-cost gas sources are higher. In Eastern Europe, comparably cheap Russian gas is abundant, however, these countries are not as flexible in their supplier choice due to infrastructure constraints. Therefore the traders present in those markets, above all the Russian trader, make higher profits than their counterparts in Western European markets. For Eastern European countries, we see in \mbox{Figure \ref{fig:P2_PriceDev}} that the traders' profits decrease over time, as infrastructural constraints diminish, until the wholesale prices reach Western European levels. 

\subsection{Social welfare analysis} \label{sec:P2_SWAnalysis}

\begin{figure}[htbp]  
\centering
\includegraphics[width=0.9\linewidth,height=0.96\textheight,keepaspectratio]{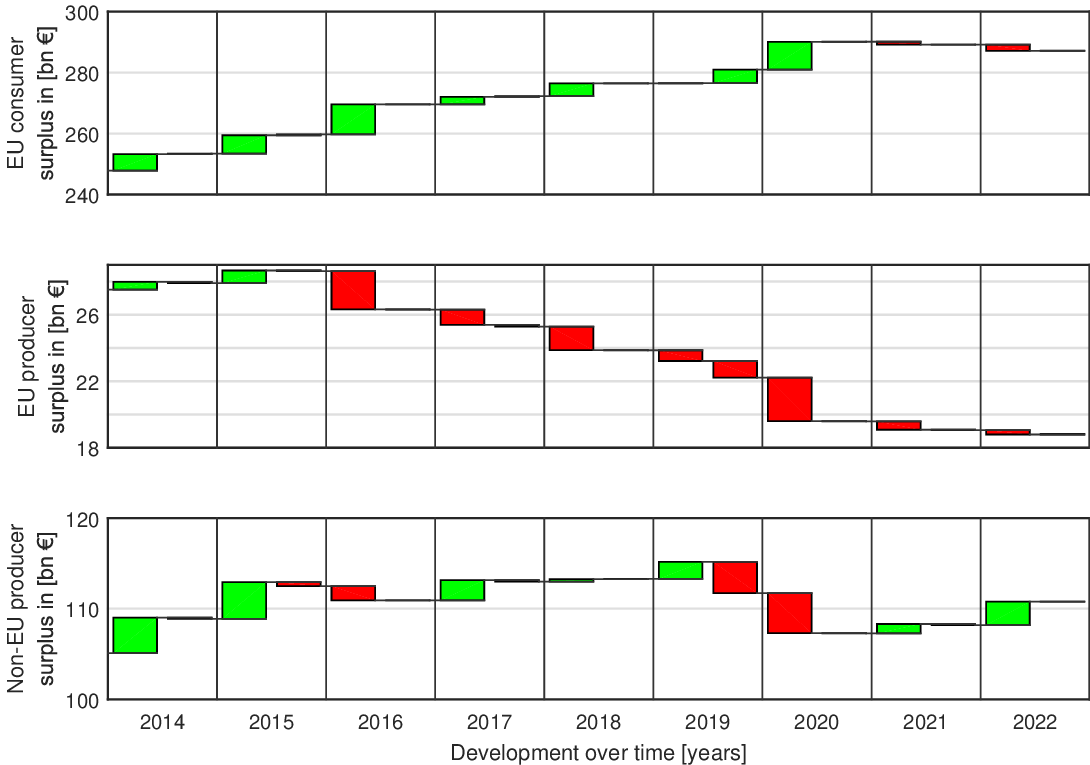}
\caption{\Sw{} development over time. Subplots 1-3 show the development of the \cs{} in the \ac{EU}, the \ps{} in the \ac{EU}, and the \ps{} of all non-\ac{EU} countries supplying the \ac{EU}. The bars on the left-hand side in every year show the effects on \cps{} caused by changes in production capacities, liquefaction capacities, demand, and \wtp{} ($\SIM_{y,3}-\SIM_{y,0}$), while the bars on the right-hand side show the effects induced by the expansions of the other infrastructure types ($\SIM_{y,k_y+4}-\SIM_{y,3}$). Green indicates a growth, and red a decline.}
\label{fig:P2_dSWEU} 
\end{figure}

As \mbox{Figure \ref{fig:P2_dSWEU}} indicates, the \cs{} in the \ac{EU} increases by \unit[17.0]{\%} from 2013 to 2020, and then decreases slightly to a total of \unit[+15.9]{\%} over the simulation horizon. The surplus of the producers within the \ac{EU} is one order of magnitude smaller than the \cs{}. After a slight increase in 2014 and 2015, it decreases by a total of $\unit[31.6]{\%}$ over the horizon. The main reason for this decrease is the declining production in the \ac{EU} (see also \mbox{Figure \ref{fig:P2_RelConsumptionByTrader}}). For producers outside the \ac{EU}, the surplus remains roughly constant over the simulation horizon.

When comparing the left-hand and right-hand side bars, we note that only in 2019 the \cs{} increases significantly from infrastructure expansions, despite the numerous expansions planned over the simulation horizon. All other changes in \cs{} are predominantly from the favorable development of production capacities in relation to demand and \wtp{}. As this is a rather unexpected result, we will analyze the situation in 2019 in detail in the remainder of this section.

\begin{figure*}[!htb]
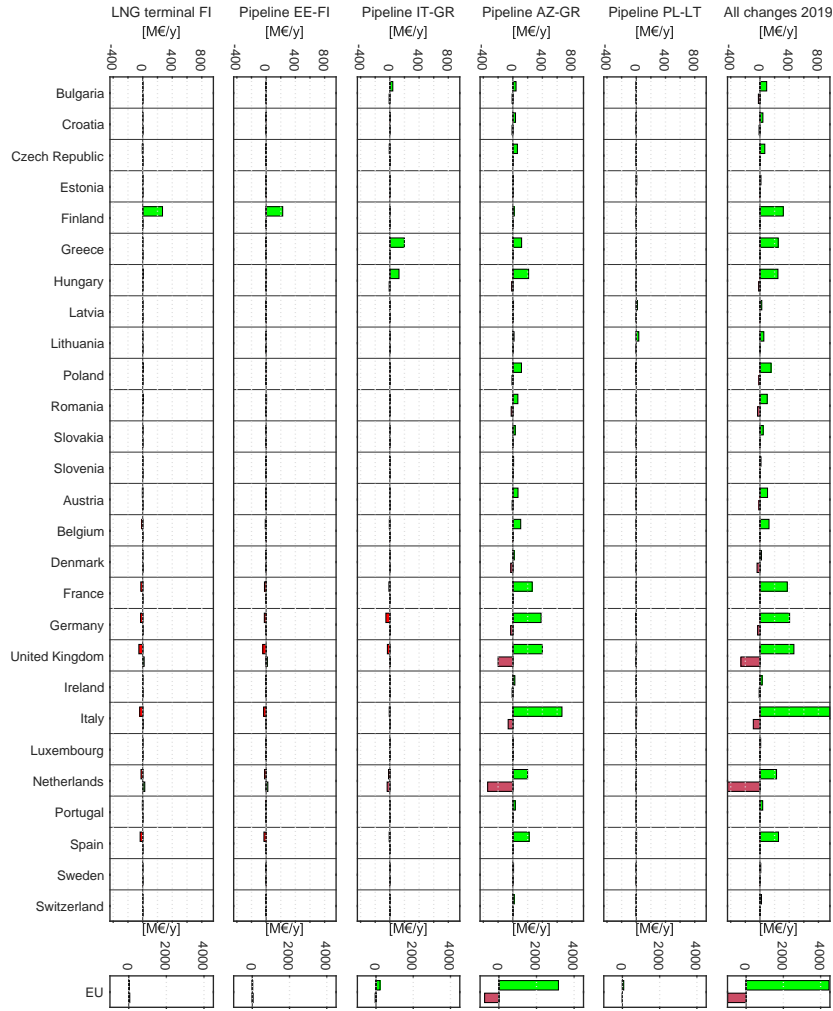

\caption{Changes in \cps{} in the \ac{EU}-member states and in the \ac{EU} as a whole from infrastructure expansions in 2019. Results are shown for five individual infrastructure expansions ($\SIM_{2019,\ID}$, for $\ID \in \{7,15,20,22,24\}$), and for all changes taking place together ($\SIM_{2019,28}$, corresponding to the infrastructure state at the end of 2019). For each country, the top bar represents the change in \cs{}, while the bottom bar indicates the change in \ps{}, compared to $\SIM_{2019,3}$. Abbreviations: FI: \FI{}. EE: \EE{}. IT: \IT{}. GR: \GR{}. AZ: \AZ{}. PL: \PL{}. LT: \LT{}.}
\label{fig:P2_dCSdPSPerCountry} 
\end{figure*}

\mbox{Table \ref{tab:P2_ImportantInfrExp2019}} lists all infrastructure expansions planned to come on-line during 2019 and their impact on \sw{}. The most influential projects in terms of \sw{} changes summed over all \ac{EU} countries are shown in italics, and their consequences on \cps{} detailed in \mbox{Figure \ref{fig:P2_dCSdPSPerCountry}}.  Due to complementary or substitution effects of the projects, the sum of their individual impacts on \sw{} does not match the total \sw{} change at the end of 2019. Nevertheless, analyzing individual projects hints at the geographical region they influence and the magnitude of their impact. 

{\small \singlespacing \renewcommand\arraystretch{1.5}
\addtolength{\tabcolsep}{-4pt}
\begin{longtabu}  to 0.9\linewidth {llrrr}
\caption{Impacts of infrastructure expansions and changes in demand on the market for the year 2019, listed by simulation $\ID$, where $\SIM_{2019,\ID}$ is the corresponding simulation. The changes in \ac{EU} \sw{} in $\SIM_{2019,\ID}$, $\ID \in \{1,2,3\}$ are relative to $\SIM_{2019,0}$, while all other changes are relative to $\SIM_{2019,3}$ (cf.\, \mbox{Section \ref{sec:P2_SimulationRuns}}). The regasification, storage, and pipeline expansions with highest impact on \sw{} in the \ac{EU} are highlighted in italics. Further details on those simulations are provided in \mbox{Figure \ref{fig:P2_dCSdPSPerCountry}}. \unitfrac{M\euro{}}{y}: Million Euros per year.}
\label{tab:P2_ImportantInfrExp2019}\\
\toprule 
$\boldsymbol{\ID}$ 	& \textbf{Infrastructure expansion}	& \multicolumn{1}{c}{\textbf{Original}} 	& \multicolumn{1}{c}{\textbf{Capacity}} & \multicolumn{1}{c}{\textbf{Total \ac{EU}}} 	\\ 
								&																		&  \multicolumn{1}{c}{\textbf{capacity}	}	&  \multicolumn{1}{c}{\textbf{change}}	& \multicolumn{1}{c}{\textbf{social wel-}} \\ 
								&																		&  																				&  																			& \multicolumn{1}{c}{\textbf{fare change}} \\ 
\midrule
\endfirsthead
\multicolumn{5}{c}{\tablename\ \thetable\ -- \textit{Continued from previous page}} \\
\toprule
$\boldsymbol{\ID}$ 	& \textbf{Infrastructure expansion}	& \multicolumn{1}{c}{\textbf{Original}} 	& \multicolumn{1}{c}{\textbf{Capacity}} & \multicolumn{1}{c}{\textbf{Total \ac{EU}}} 	\\ 
								&																		&  \multicolumn{1}{c}{\textbf{capacity}	}	&  \multicolumn{1}{c}{\textbf{change}}	& \multicolumn{1}{c}{\textbf{social wel-}} \\ 
								&																		&  																				&  																			& \multicolumn{1}{c}{\textbf{fare change}} \\ 
\midrule
\endhead
\bottomrule 
\multicolumn{5}{r}{\textit{Continued on next page}} \\
\endfoot
\bottomrule
\endlastfoot
\multicolumn{4}{l}{\textbf{Production, liquefaction, and demand, relative to $\SIM_{2019,0}$}}									&	\\*
1 & \multicolumn{3}{l}{Production and liquefaction capacity changes in all countries}				& \unitfrac[-1015]{M\euro{}}{y}	\\ 
2 &\multicolumn{3}{l}{Demand and \wtp{} changes in all countries}									& \unitfrac[606]{M\euro{}}{y}	\\ 
3 & \multicolumn{3}{l}{Combined effects from $\SIM_{2019,1}$ and $\SIM_{2019,2}$}	& \unitfrac[-587]{M\euro{}}{y}		\\
\midrule
\multicolumn{2}{l}{\textbf{Regasification terminals}}															&													&		&	\\*
4 & \ES{}				 																& \unitfrac[206]{mcm}{d}		& \unitfrac[5]{mcm}{d}		& \unitfrac[0]{M\euro{}}{y}	\\ 
5 & \FR{}				 																& \unitfrac[144]{mcm}{d}		& \unitfrac[5]{mcm}{d}		& \unitfrac[0]{M\euro{}}{y}	\\ 
6 & \IT{}				 																& \unitfrac[172]{mcm}{d}		& \unitfrac[24]{mcm}{d}		& \unitfrac[0]{M\euro{}}{y}	\\ 
\textit{7} & \textit{\FI{}}				 							& \textit{\unitfrac[0]{mcm}{d}}		& \textit{\unitfrac[19]{mcm}{d}}& \textit{\unitfrac[+87]{M\euro{}}{y}}	\\
\midrule
\multicolumn{2}{l}{\textbf{Storage facilities}} && &\\*
8 & \ES{}, extraction				 										& \unitfrac[60]{mcm}{d} 		& \unitfrac[1]{mcm}{d}		& \unitfrac[0]{M\euro{}}{y}	\\
9 & \DE{}, injection				 										& \unitfrac[212]{mcm}{d} 		& \unitfrac[13]{mcm}{d}		& 	\\
	&\DE{}, extraction				 										& \unitfrac[541]{mcm}{d} 		& \unitfrac[13]{mcm}{d}		& \\
	&\DE{}, capacity				 											& \unit[22525]{mcm} 				& \unit[120]{mcm}					& \unitfrac[0]{M\euro{}}{y}\\
10 & \CZ{}, extraction				 									& \unitfrac[85]{mcm}{d} 		& \unitfrac[1]{mcm}{d}		& \unitfrac[0]{M\euro{}}{y}\\
\midrule
\multicolumn{2}{l}{\textbf{Pipelines}} && &\\*
11 & \AT{} $\rightarrow$ \DE{}							 		& \unitfrac[30]{mcm}{d}			& \unitfrac[6]{mcm}{d}		& \unitfrac[0]{M\euro{}}{y}\\
12 & \AT{} $\rightarrow$ \IT{}							 		& \unitfrac[104]{mcm}{d}		& \unitfrac[29]{mcm}{d}		& \unitfrac[0]{M\euro{}}{y}\\ 
13 & \CH{} $\rightarrow$ \FR{}							 		& \unitfrac[0]{mcm}{d}			& \unitfrac[9]{mcm}{d}		& \unitfrac[2]{M\euro{}}{y}\\ 
14 & \DE{} $\rightarrow$ \AT{}							 		& \unitfrac[38]{mcm}{d}			& \unitfrac[26]{mcm}{d}		& \unitfrac[0]{M\euro{}}{y}\\ 
\textit{15} & \textit{\EE{} $\rightarrow$ \FI{}}& \textit{\unitfrac[0]{mcm}{d}}	& \textit{\unitfrac[7]{mcm}{d}}	& \textit{\unitfrac[+72]{M\euro{}}{y}}\\ 
16 & \FI{} $\rightarrow$ \EE{}							 		& \unitfrac[0]{mcm}{d}			& \unitfrac[7]{mcm}{d}		& \unitfrac[0]{M\euro{}}{y}\\ 
17 & \FR{} $\rightarrow$ \LU{}							 		& \unitfrac[0]{mcm}{d}			& \unitfrac[4]{mcm}{d}		& \unitfrac[+1]{M\euro{}}{y}\\ 
18 & \GR{} $\rightarrow$ \IT{}							 		& \unitfrac[0]{mcm}{d}			& \unitfrac[61]{mcm}{d}		& \unitfrac[0]{M\euro{}}{y}\\ 
19 & \IT{} $\rightarrow$ \AT{}							 		& \unitfrac[18]{mcm}{d}			& \unitfrac[6]{mcm}{d}		& \unitfrac[-23]{M\euro{}}{y}\\ 
\textit{20} & \textit{\IT{} $\rightarrow$ \GR{}}& \textit{\unitfrac[0]{mcm}{d}}		& \textit{\unitfrac[48]{mcm}{d}}	& \textit{\unitfrac[+170]{M\euro{}}{y}}\\ 
21 & \LT{} $\rightarrow$ \PL{} 									& \unitfrac[0]{mcm}{d}			& \unitfrac[3]{mcm}{d}		& \unitfrac[0]{M\euro{}}{y}\\ 
\textit{22} & \textit{\AZ{} $\rightarrow$ \GR{}}& \textit{\unitfrac[0]{mcm}{d}}		& \textit{\unitfrac[66]{mcm}{d}}	&\textit{ \unitfrac[+2409]{M\euro{}}{y}}\\ 
23 & \TNL{} $\rightarrow$ \DE{}							 		& \unitfrac[143]{mcm}{d}		& \unitfrac[6]{mcm}{d}		& \unitfrac[0]{M\euro{}}{y}\\ 
\textit{24} & \textit{\PL{} $\rightarrow$ \LT{}}& \textit{\unitfrac[0]{mcm}{d}}			& \textit{\unitfrac[6]{mcm}{d}}		& \textit{\unitfrac[+65]{M\euro{}}{y}}\\ 
25 & \PL{} $\rightarrow$ \SK{}							 		& \unitfrac[0]{mcm}{d}			& \unitfrac[12]{mcm}{d}		& \unitfrac[-1]{M\euro{}}{y}\\ 
26 & \SI{} $\rightarrow$ \IT{}							 		& \unitfrac[27]{mcm}{d}			& \unitfrac[2]{mcm}{d}		& \unitfrac[0]{M\euro{}}{y}\\ 
27 & \SK{} $\rightarrow$ \PL{}							 		& \unitfrac[0]{mcm}{d}			& \unitfrac[16]{mcm}{d}		& \unitfrac[0]{M\euro{}}{y}\\
\midrule
28 & \multicolumn{3}{l}{Effect of the combination of all expansions 4-27}															& \unitfrac[+3456]{M\euro{}}{y}\\*
\end{longtabu}
}

\begin{figure*}[!htb]
\centering
\includegraphics[height=0.9\textwidth,width=0.96\textheight,keepaspectratio,angle=-90]{025dcap_112dCSdPSPerCountry_0.eps}
\end{figure*}

As \mbox{Table \ref{tab:P2_ImportantInfrExp2019}} shows, the new connection from \AZ{} to \GR{}, also known as \ac{TANAP}, contributes significantly to the overall changes in \sw{} in 2019 ($\SIM_{2019,22}$). By comparing ``Pipeline AZ-GR'' to ``All changes 2019'' in \mbox{Figure \ref{fig:P2_dCSdPSPerCountry}}, we also notice that most of the effects on \sw{} are similar, which suggests that they originate from the \ac{TANAP}. The \ac{TANAP} increases \cs{} between \unitfrac[5]{M\euro}{y} (\EE{}) and \unitfrac[665]{M\euro}{y} (\IT{}), which corresponds to an increase of \unit[1-2]{\%} for most countries, and \unit[5.2]{\%} in \GR{} (\mbox{\ref{app:P2_ExcelTable}}). This leads to a massive boost in \cs{} of 3.2 billion Euros per year in the \ac{EU} as a whole, while \ps{} in the \ac{EU} decreases by \unitfrac[0.8]{B\euro{}}{year}. The second important project in Southeastern Europe is the \ac{TAP} connecting \GR{} and \IT{}. As shown in \mbox{Figure \ref{fig:P2_dCSdPSPerCountry}} mainly \GR{} and \HU{} can benefit from this, while most other countries see a slight decrease in \cs{}. However, when we analyze the situation at the end of 2019, we see that the joint installation of all infrastructure planned for 2019 boosts \cs{} in \IT{} to a higher level. This is because in $\SIM_{2019,28}$ the \ac{TAP} and the \ac{TANAP} are both active, allowing Azeri gas to be traded all over Europe, and particularly in \IT{}. This illustrates well how the market impact of two complementary pipelines can be underestimated when assessing them individually. Also note the drop of the wholesale price and the traders' profits in \GR{} induced by these infrastructure expansions (\mbox{Figure \ref{fig:P2_PriceDev}}). 

The second region where projects with high impact are implemented is Northeastern Europe: a regasification terminal is installed in \FI{}, and two pipelines connecting \EE{} with \FI{}, and \PL{} with \LT{} are commissioned. As a consequence, we see prices drop between \unitfrac[68]{k\euro{}}{mcm} (\FI{}) and \unitfrac[11]{k\euro{}}{mcm} (\EE{}) compared to 2018 (\mbox{Figure \ref{fig:P2_PriceDev}}), and \cs{} change by \unitfrac[+437]{M\euro{}}{y} or \unit[+13]{\%} (thereof \unitfrac[+337]{M\euro{}}{y} in \FI{} corresponding to a \unit[25]{\%} increase; percentages are calculated from \mbox{\ref{app:P2_ExcelTable}}). This is because the additional infrastructure allows British, Dutch and Norwegian gas to be traded in the region (\mbox{Figure \ref{fig:P2_RelConsumptionByTrader}}). This increases average gas production costs by more than \unit[50]{\%}, but also significantly lowers the traders' profits (\mbox{Figure \ref{fig:P2_PriceDev}}). Overall, prices in the Baltic region move close to Western European levels, indicating that the region is well interconnected to the rest of the market after 2019.

As \mbox{Figure \ref{fig:P2_dCSdPSPerCountry}} shows, \ps{} falls significantly for all \ac{EU} producers. The volumes brought to the \ac{EU} markets by \AZ{} induce a high pressure on prices and market shares of the other traders, and as a consequence, their surpluses decrease. However, we can observe from \mbox{Figure \ref{fig:P2_dSWEU}} that \cps{} do not always move in opposite directions. For instance in 2014 and 2015, \cps{} in- and outside the \ac{EU} grow, because the changes in demand, \wtp{}, and production capacities are in favor of both, the consumers and the producers, and therefore benefits are shared among them.

\subsection{Categorization of projects} \label{sec:P2_CatOfProjects}
The results presented above allow us to classify the infrastructure projects foreseen for the coming decade into three categories. Note that there is some uncertainty inherent to this categorization as we only assessed projects individually, and only at the point in time they are commissioned. The first category contains projects significantly boosting \sw{} in the \ac{EU} as a whole. The only project fulfilling this criterion is the \ac{TANAP}. This project is unique since it makes large quantities accessible for the \ac{EU} market at a comparably low price. The consequences can be felt all over Europe as the changes in \cps{} are significant in most countries.

A second group of projects does not add significant amounts of gas to the \ac{EU} market, but instead changes the distribution of gas within the market. Examples include the ones presented in \mbox{Figure \ref{fig:P2_dCSdPSPerCountry}}. As depicted, these projects mainly benefit the destination regions, while most other countries lose \cps{}, which lowers the overall \sw{} gains \ac{EU}-wide, or even turns them negative as illustrated by the examples of $\SIM_{2019,19}$ (\mbox{Table \ref{tab:P2_ImportantInfrExp2019}}).

For a third surprisingly large group of projects the effects on the market are marginal. As \mbox{Table \ref{tab:P2_ImportantInfrExp2019}} indicates, this is the case for a majority of the projects in 2019. Overall, 130 out of 171 projects move \sw{} in the \ac{EU} by less than \unitfrac[10]{M\euro{}}{y} (\mbox{\ref{app:P2_ExcelTable}}). As most of the storage expansion projects fall into this category we can conclude that there is enough storage capacity available to cover seasonal demand fluctuations and there is no need for additional storage capacity in most regions \textendash at least in the case of a fully functional market. These results coincide with the findings of \citet[Chapter 6.2.3]{ENTSO-G2015} estimating a storage utilization rate of less than \unit[50]{\%}. 
A notable project in this category is the recently announced Turkish Stream: although the pipeline will allow Russian gas to enter the \ac{EU} from a different angle (and therefore potentially increase supplier diversification for countries not yet supplied by Russian gas), there are two major obstacles to overcome: first, the infrastructure within the \ac{EU} to bring the additional gas to the larger markets in Western Europe is lacking. The planned infrastructure in Southeastern Europe, including the \ac{TAP}, will already be utilized by the gas arriving from \AZ{}, and hence, Russian gas can at most substitute Azeri gas, but cannot increase the shipped quantity overall; and second, there are shorter and thus cheaper routes to ship Russian gas to the major consumer markets, namely, via \BY{} and the \UA{}. Hence, it is hard to think of situations in a functioning market in which the Turkish Stream will be the first choice for shipping Russian gas to the \ac{EU}. As the capacity limit of the Turkish Stream pipeline as it is implemented in our model is not reached by 2022, it is safe to assume that a less restrictive assumption on the pipeline's availability and capacity (\mbox{Section \ref{sec:P2_TransportationModeling}}) does not change the results. 

\section{Conclusions and Outlook} \label{sec:P2_Conclusions}

We analyzed the gas price and consumption development from 2013 to 2022 resulting from changes in demand, production capacity, and infrastructure. We found that the wholesale prices drop on average over time in all \ac{EU} markets, contradicting the trend predicted by the \citet{Iea2013}. Furthermore, we observed that the price levels in the \ac{EU} have largely converged by 2019. Thus, we envision that the \ac{EU} market's physical integration is completed with the connection of \FI{} to the \ac{ENTSO-G} network in 2019 and, unless there are significant regional supply shortages or demand increases, there is no reason to expect prices to spike regionally thereafter.

The benefits of integration are clearly greater in Eastern than in Western Europe: prices fall more significantly as the newly built infrastructure introduces new possibilities for arbitrage, which in turn reduces the traders' profits. This stimulates consumption and \cs{}. Furthermore, in the Baltic states and \FI{}, the most exposed countries in 2013, the \ac{HHI} is halved by 2019, indicating a significant increase in supplier diversity. Some concerns remain regarding the development of the supplier diversity in \PL{}, \SK{}, and \RO{}, because these countries import most of their (steadily increasing) needs from a single supplier throughout the entire simulation horizon.

From our simulation results, we identified three categories of projects regarding \cps{} changes. A first category comprises infrastructure expansions bringing additional gas to the \ac{EU} markets and thereby increasing \cs{} significantly. The only project in this group is the \ac{TANAP}, which brings gas from \AZ{} to the \ac{EU} border in \GR{}. The project is part of the ``Southern corridor'' and is brought on-line in 2019.

The second group comprises projects which are highly beneficial for a single country or region, but have minimal or even negative impact on the \sw{} of other European countries. We provided several examples, including the \ac{LNG} terminal in \FI{} or the \ac{TAP} connecting \IT{} and \GR{}. These projects strengthen the market integration, and thereby lower prices and induce consumption in the destination country. The other countries, however, face additional competition for the existing gas volumes and as a consequence experience higher prices and a drop in \sw{}.

A third category of projects has only a marginal impact on the market. A possible explanation for the existence of this group is that these infrastructures are built for situations which were not analyzed in our study, such as emergency situations. Another explanations is that the infrastructures are used for activities which are not captured by our model: The pipelines and regasification terminals could be used by the consumers to diversify suppliers beyond the levels assumed in our study, for instance, due to a high perceived risk of disruptions. A high risk perception could also fuel seasonal gas storage usage. Furthermore, injection and extraction capacities are potentially expanded for peak shaving purposes, an effect which is not captured by our model. Either way, the purpose of the projects which are not utilized under normal operation should be investigated in a future study.

Finally, we found that most of the big shifts in \sw{} originate from changes in production capacities, demand, and \wtp{}, and not from infrastructure expansions. Consequently, it may be advisable for the \ac{EU} to focus on these aspects, especially after the completion of the internal market in 2019. At the same time, a larger total \sw{} is not necessarily beneficial for all parties involved in gas trade: it largely matters whether consumer or \ps{} changes, and in which countries these values change. Furthermore, the situation for consumers can also be improved by efficiency measures, which leads to a decrease of \sw{}. Future work should elaborate on possible development paths of the supply-to-demand ratio and assess how producers and consumers are affected by various scenarios. We also envision additional research on the costs and benefits of higher production capacities and lower demand within the \ac{EU} borders. In addition, the combined effects of infrastructure elements should be assessed in more detail; our results indicate the main regions of interest.

\section*{Acknowledgements}
\addcontentsline{toc}{section}{Acknowledgements}
We would like to thank Prof.~Ruud Egging for his valuable comments.

\clearpage
\appendix
\section{Model equations and exemplary market setting}  \label{app:P2_ModelEq_Pic_Notation}
\mbox{\ref{app:P2_ModelEq_Pic_Notation}} is largely reproduced from \citet{Baltensperger2015}. We introduce the model equations in \mbox{\ref{app:P2_ModelEquationsSubSec}}, show an exemplary market setting with two interconnected nodes in \mbox{\ref{app:P2_figure}}, and introduce the associated notation in \mbox{\ref{app:P2_NotationSubSec}}. Note that we follow the convention used by \citet{Baltensperger2015} and include producers in the notion of service providers, although producers are not an infrastructure service.

\subsection{Model equations} \label{app:P2_ModelEquationsSubSec}
Equations \eqref{eqn:P2_NoLOSSnoTauNoS} describe the mechanics of the spatial partial equilibrium model of the European gas market in detail. We refrain from showing the loss terms in the equations to achieve a more compact notation. 
\begin{subequations} \label{eqn:P2_NoLOSSnoTauNoS}
\allowdisplaybreaks
\begin{alignat}{4}
0 &\leq& \, \LINC^P_{nt} + \QUAC^P_{nt} q^P_{fnt} + \alpha^P_{nt} + \alpha^{PT}_{n} - \phi^N_{fnt} & \perp q^P_{fnt} &&\,\geq 0 \quad \forall f,n,t \label{eqn:P2_dqP}\\
0 &\leq&\, \LINC^I_{nt} + \alpha^I_{nt} + \alpha^{IT}_{n} + \phi^N_{fnt} - \phi^S_{fn}& \perp q^I_{fnt} &&\,\geq 0 \quad \forall f,n,t \label{eqn:P2_dqI}\\
0 &\leq&\, \LINC^X_{zt} + \alpha^X_{nt} + \alpha^{XT}_{n} - \phi^N_{fnt} + \phi^S_{fn} & \perp q^X_{fnt} &&\,\geq 0 \quad \forall f,n,t \label{eqn:P2_dqX}\\
0 &\leq&\, \LINC^A_{nmt} + \alpha^A_{nmt} + \alpha^{AT}_{nm} - \phi^N_{fmt} + \phi^N_{fnt} & \perp q^A_{fnmt} &&\,\geq 0 \quad \forall f,n,m,t \label{eqn:P2_dqA}\\
0 &\leq&\, \LINC^L_{nt} + \alpha^L_{nt} + \alpha^{LT}_{n} + \LINC^B_{nmt} + \alpha^B_{nmt} + \alpha^{BT}_{nm} &&& \notag\\
&&\, + \LINC^R_{mt} + \alpha^R_{mt} + \alpha^{RT}_{m} - \phi^N_{fmt} +  \phi^N_{fnt} & \perp q^B_{fnmt} &&\,\geq 0 \quad \forall f,n,m,t \label{eqn:P2_dqB}\\
0 &\leq&\, -\lambda^C_{nt} -\theta^C_{fnt} \SLP^C_{nt} q^C_{fnt} + \phi^N_{fnt}& \perp q^C_{fnt} &&\,\geq 0  \quad \forall f,n,t \label{eqn:P2_dqC}\\
0 &\leq&\, q^{P}_{fnt} + q^{X}_{fnt} + \sum \limits_{m \in \mathcal{A}(n)} q_{fmnt}^{A} + \sum \limits_{m \in \mathcal{B}(n)} q_{fmnt}^{B}&&& \notag \\
&&\,- q^{I}_{fnt} -  q^{C}_{fnt} - \sum \limits_{m \in \mathcal{A}(n)} q_{fnmt}^{A} - \sum \limits_{m \in \mathcal{B}(n)} q_{fnmt}^{B} & \perp \phi^N_{fnt} &&\,\geq 0  \quad \forall f,n,t \label{eqn:P2_dphiN}\\
0 &\leq&\, \sum \limits_{t \in \mathcal{T}} q^{I}_{fnt} - \sum \limits_{t \in \mathcal{T}} q^{X}_{fnt} &\perp \phi^S_{fn} &&\,\geq 0  \quad \forall f,n\label{eqn:P2_dphiS}\\
0 &\leq&\, \overline{\CAP}^Z_{zt} - \sum \limits_{f \in \mathcal{F}(z)} q^{Z}_{fzt} & \perp \alpha^Z_{zt} &&\,\geq 0  \quad \forall z,t \label{eqn:P2_dalphaZ}\\
0 &\leq&\, \overline{\CAP}^{ZT}_{z} - \sum \limits_{t \in \mathcal{T}} \sum \limits_{f \in \mathcal{F}(z)} q^{Z}_{fzt} & \perp \alpha^{ZT}_z &&\,\geq 0  \quad \forall z \label{eqn:P2_dbetaZ}\\
0 &\leq&\, \lambda^C_{nt} - \left( \INT^C_{nt} + \SLP^C_{nt} \sum \limits_{f \in \mathcal{F}(n)} q^{C}_{fnt} \right) &\perp \lambda^C_{nt} &&\,\geq 0  \quad \forall n,t\label{eqn:P2_dpiC}
\end{alignat}
\end{subequations}

\clearpage
\subsection{Graphical illustration} \label{app:P2_figure}
\begin{figure*}[!htb]
\centering
\includegraphics[width=0.9\textwidth]{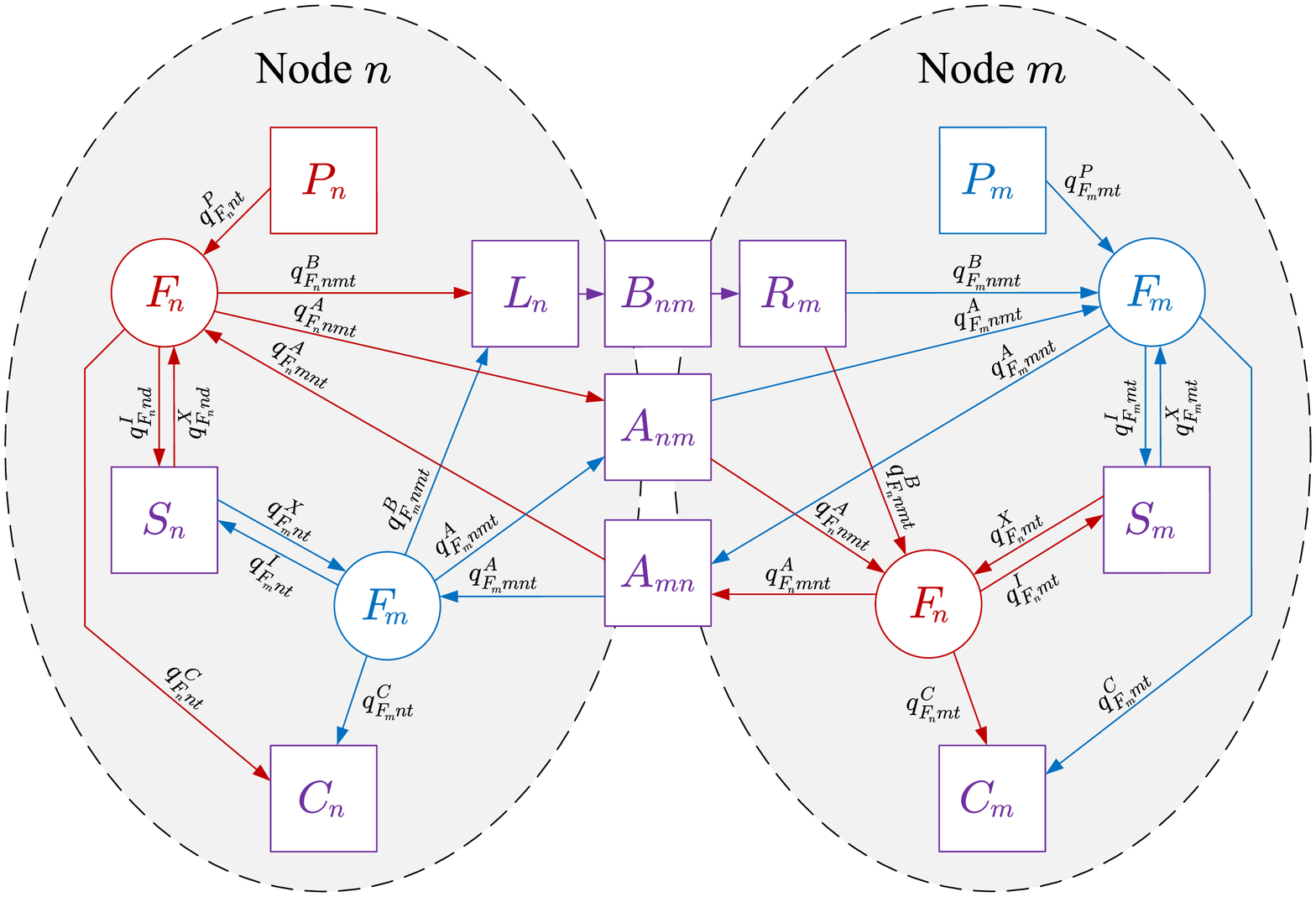}
\end{figure*}

\begin{figure*}[!htb]
\caption{Gas market model with two nodes. $P$: producer. $A$: pipeline operator. $L$: liquefaction plant operator. $B$: LNG shipment. $R$: regasification plant operator. $S$: storage operator. $C$: consumer. $F_n$: trader associated with producer $P_n$. $F_m$: trader associated with producer $P_m$. $q^{P}_{fnt}$: quantity delivered from producer to trader $f$ at node $n$ in time period $t$. $q^{A}_{fnmt}$: pipeline transportation of trader $f$ via arc $nm$ in period $t$. $q^{B}_{fnmt}$: LNG shipment of trader $f$ via arc $nm$ in period $t$. $q^{I}_{fnt}$: storage injection by trader $f$ at node $n$ in period $t$. $q^{X}_{fnt}$: storage extraction by trader $f$ at node $n$ in period $t$. $q^{C}_{fnt}$: sales by trader $f$ to consumer in node $n$ in period $t$. The traders $F_n$ and $F_m$, there decision variables, and their corresponding producers $P_n$ and $P_m$ are colored red ($n$) and blue ($m$), respectively. Service providers, except producers, and flows between them are marked purple, as well as the consumers, since all traders trade with them.}
\label{fig:P2_Model_full}
\end{figure*}

\clearpage
\subsection{Notation} \label{app:P2_NotationSubSec}

\begin{table}[htbp]
\caption{This table introduces the nomenclature concerning service providers, traders and consumers.}
\label{tab:P2_ServiceProviders}
\centering \small \renewcommand\arraystretch{1.5}
\begin{tabu} to 0.9\linewidth {lX}
\toprule 
\multicolumn{2}{l}{\textbf{Service providers, traders and consumers}} \\
\midrule 
$A_{nm}$& Transmission system operator of \mbox{pipeline $nm$}\\
$B_{nm}$& Shipping company transporting \ac{LNG} from $n$ \mbox{to $m$}\\
$C_n$& Consumer at \mbox{node $n$}\\
$I_n$& Storage operator injecting gas at \mbox{node $n$}\\
$L_n$& Liquefaction plant operator at \mbox{node $n$}\\
$P_n$& Gas producing company at \mbox{node $n$}\\
$R_n$& Regasification plant operator at \mbox{node $n$}\\
$S_n$& Storage operator at \mbox{node $n$}\\
$F_n$& The trader associated with producer at \mbox{node $n$}\\
$X_n$& Storage operator extracting gas at \mbox{node $n$}\\
$Z_z$& Placeholder for a service provider ($P_n$, $I_n$, $X_n$, $L_n$, $R_n$, $A_{nm}$, $B_{nm}$) at node $n$ / \mbox{arc $nm$}\\
\bottomrule
\end{tabu}
\end{table}

{\small \singlespacing \renewcommand\arraystretch{1.5}
\begin{longtabu}  to 0.9\linewidth {lX}
\caption{This table introduces all sets used for the mathematical description of the model.}
\label{tab:P2_Sets} \\
\toprule 
\multicolumn{2}{l}{\textbf{Sets}}  \\
\midrule 
\endfirsthead
\multicolumn{2}{c}{\begin{footnotesize}\tablename\ \thetable\ -- \textit{Continued from previous page}\end{footnotesize}} \\
\toprule
\multicolumn{2}{l}{\textbf{Sets}}  \\
\midrule
\endhead
\bottomrule
\multicolumn{2}{r}{\begin{footnotesize}\textit{Continued on next page}\end{footnotesize}} \\
\endfoot
\bottomrule
\endlastfoot
$t \in \mathcal{T} = \{T_1, \ldots, T_{\bar{t}}\}$& A time period $t$ in the set $\mathcal{T}$ of all periods of a year\\
$n, m \in \mathcal{N} = \{N_1, \ldots, N_{\bar{n}}\}$& Nodes $n,m$ in the set $\mathcal{N}$ of all nodes \\ 
$n^\ast(f) \in \mathcal{N}$& Node from which trader $f$ draws its gas \\
$f \in \mathcal{F} = \{F_1, \ldots, F_{\bar{n}}\}$& A trader $f$ in the set $\mathcal{F}$ of all traders\\
$z \in \mathcal{Z}$ & A node/arc element from the \mbox{set $\mathcal{Z}$} \\
\midrule
$\mathcal{A} \subset \mathcal{N} \times \mathcal{N}$& Set of arcs connecting $2$ nodes by pipeline					\\
$\mathcal{B} \subset \mathcal{N} \times \mathcal{N}$ & Set of arcs connecting $2$ nodes by ship					\\	
$\mathcal{C} \subseteq \mathcal{N}$ & Set of nodes at which a consumer is active \\
$\mathcal{I} \subseteq \mathcal{N}$ & Set of nodes at which storage injection	is possible					\\
$\mathcal{L} \subseteq \mathcal{N}$ & Set of nodes at which a liquefaction terminal operator is active				\\	
$\mathcal{P} \subseteq \mathcal{N}$ & Set of nodes at which a gas producer is active	\\							
$\mathcal{R} \subseteq \mathcal{N}$ & Set of nodes at which a regasification terminal operator is active		\\
$\mathcal{X} \subseteq \mathcal{N}$ & Set of nodes at which storage extraction is possible						\\
$\mathcal{Z} \in \{\mathcal{P},\mathcal{L},\mathcal{B},\mathcal{R},\mathcal{A},\mathcal{I},\mathcal{X}\}$ & Placeholder for the set of nodes/arcs at which a type of service provider is active \\
\midrule
$\mathcal{A}(n) \subseteq \mathcal{N} \setminus \{n\}$ & Set of nodes which are connected to $n$ by pipeline \\
$\mathcal{B}(n) \subseteq \mathcal{N} \setminus \{n\}$ & Set of nodes which are connected to $n$ by ship \\
$\mathcal{C}(f) \subseteq \mathcal{N}$& The set of all nodes with consumers which are reachable by \mbox{trader $f$}\\
$\mathcal{N}(f) \subseteq \mathcal{N}$& The set of all nodes which are reachable by \mbox{trader $f$}\\
$\mathcal{F}(z)$& The set of all traders active at node/\mbox{arc $z$}\\
$\mathcal{Z}(f)$& The set of all nodes/arcs in which service $Z$ is active and are reachable by \mbox{trader $f$}\\ 
\end{longtabu}
}

{\small \singlespacing \renewcommand\arraystretch{1.5}
\begin{longtabu}  to 0.9\linewidth {lX}
\caption{The \textit{parameters} are generally described by capital Roman letters. Lower-case Roman letters are only chosen if the parameter is directly linked to a variable of the same name. Occasionally, lower-case Greek letters are chosen to follow conventions. The superscripts indicate whether the parameter is related to a service provider of type $Z \in \{P,L,B,R,A,I,X\}$ or a consumer $C$. Subscripts indicate the trader $f$, node/arc $z$, and the period of the year $t$ the parameter is related to.}
\label{tab:P2_Parameters} \\
\toprule 
\multicolumn{2}{l}{\textbf{Parameters}}  \\
\midrule 
\endfirsthead
\multicolumn{2}{c}{\begin{footnotesize}\tablename\ \thetable\ -- \textit{Continued from previous page}\end{footnotesize}} \\
\toprule
\multicolumn{2}{l}{\textbf{Parameters}}  \\
\midrule
\endhead
\bottomrule
\multicolumn{2}{r}{{\footnotesize\textit{Continued on next page}}} \\
\endfoot
\bottomrule
\endlastfoot
$\overline{\CAP}^Z_{nt}$		& Maximum capacity of service $Z$ located at $z$ in \mbox{period $t$}\\
$\overline{\CAP}^{ZT}_{n}$	& Maximum capacity of service $Z$ located at $z$ over all \mbox{periods $\mathcal{T}$}\\
$\etaCcalib_{nt}$ 					& Calibrated price elasticity in market $n$ and time \mbox{period $t$}\\
$\etaCdata_{nt}$ 						& Aggregated price elasticity from data for node $n$ and time \mbox{period $t$}\\
$\INT^C_{nt}$ 							& Maximum willingness to pay (intercept of inverse demand function with the $s^C_{nt}=0$ - axis) of consumers at node $n$ in \mbox{period $t$} \\
$\piCcalib_{nt}$						& Calibrated \wtp{} for $\sCcalib_{nt}$ \\
$\piCdata_{nt}$							& Reported wholesale price levels\\
$\LINC^Z_{zt}$ 							& Linear cost function term for service $Z$ located at $z$ in \mbox{period	$t$}\\
$\LOSS^Z_z $								& Loss factor when using service $Z$  located \mbox{at $z$} \\
$\qCdata_{fnt} $						& Reference gas sales of trader $f$ in node $n$ and time \mbox{period $t$}\\
$\qCdataApprox$ 						& Adjusted reference gas sales  of trader $f$ in node $n$ and time period $t$ for calibration \\
$\QUAC^Z_{zt}$ 							& Quadratic cost function term for service $Z$ located at $z$ in \mbox{period $t$}\\
$\sCdata_{nt}$							& Reference consumption levels\\
$\sCcalib_{nt}$							& Calibrated consumption levels\\
$\SLP^C_{nt}$								& Slope of the inverse demand curve of the consumers at node $n$ in \mbox{period $t$}, is assumed strictly negative\\
$\MP_{fnt}$									& Market power parameter of trader $f$ in node $n$ and \mbox{period $t$}\\ 
\end{longtabu}
}

\begin{table}[htbp]
\caption{The \textit{variables} are described by lowercase letters. Primal variables are Roman, while dual variables are Greek letters. The superscripts indicate whether the variable is related to a service provider of type $Z \in \{P,L,B,R,A,I,X\}$, a consumer $C$, or a node $N$. Subscripts indicate the trader $f$ the variable corresponds to, at which node/arc $z$ the transaction or service is located, and in which period of the year $t$ it takes place.}
\label{tab:P2_Variables}
\centering \small \renewcommand\arraystretch{1.5}
\begin{tabu} to 0.9\linewidth {lX}
\toprule
\multicolumn{2}{l}{\textbf{Variables}} \\
\midrule
$q^{C}_{fnt}$ & Flow of trader $f$ to consumer $C$ at node $n$ in \mbox{period $t$}\\
$q^{Z}_{fzt}$ & Flow between trader $f$ and service provider $Z$ at node/arc $z$ in \mbox{period $t$} \\
$s^{C}_{nt}$ & Consumption in node $n$ and \mbox{period $t$} \\
$s^{Z}_{zt}$ & Volume flow contracted by service provider $Z$ at node/arc $z$ in \mbox{period $t$} \\
$\alpha^Z_{zt}$ & Congestion fee of service $Z$ located at $z$ in \mbox{period $t$}\\
$\alpha^{ZT}_{z}$ & Congestion fee on annual usage of service $Z$ located at $z$ \\
$\phi^N_{fnt}$ & Dual variable of the volume balance of trader $f$ at node $n$ and \mbox{period $t$}\\
$\phi^S_{fn}$ & Dual variable of the annual volume balance of trader $f$ in storage $S$ at \mbox{node $n$}\\
$\lambda^C_{nt}$ & Wholesale price at node $n$ in \mbox{period $t$}\\
\bottomrule
\end{tabu}
\end{table}

\begin{table}[htbp]
\caption{This table introduces the \textit{functions}. The superscripts indicate whether the function is related to a service provider of type $Z \in \{P,L,B,R,A,I,X\}$, or a consumer $C$. Subscripts indicate at which node $n$ the service/consumer is located, and in which period of the year $t$ the function is valid.}
\label{tab:P2_Functions}
\centering \small \renewcommand\arraystretch{1.5}
\begin{tabu} to 0.9\linewidth {lX}
\toprule 
\multicolumn{2}{l}{\textbf{Functions}} \\
\midrule 
$c^{Z}_{zt}(s^Z_{zt})$ & Cost function of service $Z$ at node/arc $z$ in \mbox{period $t$}.\\
$\Lambda^{C}_{nt}(s^C_{nt})$ & Inverse demand function of consumer $C$ at node $n$ in \mbox{period $t$}.\\
\bottomrule
\end{tabu}
\end{table}

\clearpage
\section{Calibration results} \label{app:P2_allCalibParameters}

\begin{table}[htbp]
\caption{Mean market power parameter values for all traders (T) in all markets (C). Country abbreviations are given in \mbox{Table \ref{tab:P2_CountryAbbr}}.}
\label{tab:P2_CVvalues}
\centering \small \renewcommand\arraystretch{1.5}
\addtolength{\tabcolsep}{-5pt}
\begin{scriptsize}
\begin{tabu} to 0.9\linewidth {crrrrrrrrrrrrrrrrr}
    \toprule
\backslashbox{C}{T} & AZ    & DK    & DZ$^L$    & DZ$^N$   & EG    & GB    & LY    & NG    & NL    & NO$^L$    & NO$^N$   & OM    & PE    & QA    & RU    & TT    & YE \\
    \midrule
    BG    & \ValuesColored{88}0.44 & \ValuesColored{138}0.69 & \ValuesColored{190}0.95 & \ValuesColored{162}0.81 & \ValuesColored{88}0.44 & \ValuesColored{166}0.83 & \ValuesColored{168}0.84 & \ValuesColored{200}1 & \ValuesColored{158}0.79 & \ValuesColored{88}0.44 & \ValuesColored{156}0.78 & \ValuesColored{88}0.44 & \ValuesColored{88}0.44 & \ValuesColored{184}0.92 & \ValuesColored{50}0.25 & \ValuesColored{88}0.44 & \ValuesColored{88}0.44 \\
    CZ    & \ValuesColored{38}0.19 & \ValuesColored{0}0 & \ValuesColored{0}0 & \ValuesColored{8}0.04 & \ValuesColored{0}0 & \ValuesColored{20}0.1 & \ValuesColored{36}0.18 & \ValuesColored{0}0 & \ValuesColored{26}0.13 & \ValuesColored{0}0 & \ValuesColored{30}0.15 & \ValuesColored{0}0 & \ValuesColored{0}0 & \ValuesColored{68}0.34 & \ValuesColored{72}0.36 & \ValuesColored{0}0 & \ValuesColored{0}0 \\
    EE    & \ValuesColored{72}0.36 & \ValuesColored{72}0.36 & \ValuesColored{72}0.36 & \ValuesColored{72}0.36 & \ValuesColored{72}0.36 & \ValuesColored{72}0.36 & \ValuesColored{72}0.36 & \ValuesColored{72}0.36 & \ValuesColored{72}0.36 & \ValuesColored{72}0.36 & \ValuesColored{72}0.36 & \ValuesColored{72}0.36 & \ValuesColored{72}0.36 & \ValuesColored{72}0.36 & \ValuesColored{72}0.36 & \ValuesColored{72}0.36 & \ValuesColored{72}0.36 \\
    FI    & \ValuesColored{80}0.4 & \ValuesColored{80}0.4 & \ValuesColored{80}0.4 & \ValuesColored{80}0.4 & \ValuesColored{80}0.4 & \ValuesColored{80}0.4 & \ValuesColored{80}0.4 & \ValuesColored{80}0.4 & \ValuesColored{80}0.4 & \ValuesColored{80}0.4 & \ValuesColored{80}0.4 & \ValuesColored{80}0.4 & \ValuesColored{80}0.4 & \ValuesColored{80}0.4 & \ValuesColored{80}0.4 & \ValuesColored{80}0.4 & \ValuesColored{80}0.4 \\
    GR    & \ValuesColored{160}0.8 & \ValuesColored{160}0.8 & \ValuesColored{160}0.8 & \ValuesColored{166}0.83 & \ValuesColored{152}0.76 & \ValuesColored{170}0.85 & \ValuesColored{160}0.8 & \ValuesColored{160}0.8 & \ValuesColored{92}0.46 & \ValuesColored{188}0.94 & \ValuesColored{120}0.6 & \ValuesColored{152}0.76 & \ValuesColored{128}0.64 & \ValuesColored{160}0.8 & \ValuesColored{174}0.87 & \ValuesColored{158}0.79 & \ValuesColored{152}0.76 \\
    HR    & \ValuesColored{124}0.62 & \ValuesColored{180}0.9 & \ValuesColored{70}0.35 & \ValuesColored{128}0.64 & \ValuesColored{124}0.62 & \ValuesColored{126}0.63 & \ValuesColored{128}0.64 & \ValuesColored{130}0.65 & \ValuesColored{128}0.64 & \ValuesColored{188}0.94 & \ValuesColored{128}0.64 & \ValuesColored{124}0.62 & \ValuesColored{66}0.33 & \ValuesColored{124}0.62 & \ValuesColored{132}0.66 & \ValuesColored{136}0.68 & \ValuesColored{124}0.62 \\
    HU    & \ValuesColored{66}0.33 & \ValuesColored{46}0.23 & \ValuesColored{46}0.23 & \ValuesColored{82}0.41 & \ValuesColored{46}0.23 & \ValuesColored{80}0.4 & \ValuesColored{74}0.37 & \ValuesColored{46}0.23 & \ValuesColored{90}0.45 & \ValuesColored{46}0.23 & \ValuesColored{96}0.48 & \ValuesColored{46}0.23 & \ValuesColored{46}0.23 & \ValuesColored{0}0 & \ValuesColored{62}0.31 & \ValuesColored{46}0.23 & \ValuesColored{46}0.23 \\
    LT    & \ValuesColored{90}0.45 & \ValuesColored{90}0.45 & \ValuesColored{90}0.45 & \ValuesColored{90}0.45 & \ValuesColored{90}0.45 & \ValuesColored{90}0.45 & \ValuesColored{90}0.45 & \ValuesColored{90}0.45 & \ValuesColored{90}0.45 & \ValuesColored{90}0.45 & \ValuesColored{90}0.45 & \ValuesColored{90}0.45 & \ValuesColored{90}0.45 & \ValuesColored{90}0.45 & \ValuesColored{90}0.45 & \ValuesColored{90}0.45 & \ValuesColored{90}0.45 \\
    LV    & \ValuesColored{62}0.31 & \ValuesColored{62}0.31 & \ValuesColored{62}0.31 & \ValuesColored{62}0.31 & \ValuesColored{62}0.31 & \ValuesColored{62}0.31 & \ValuesColored{62}0.31 & \ValuesColored{62}0.31 & \ValuesColored{62}0.31 & \ValuesColored{62}0.31 & \ValuesColored{62}0.31 & \ValuesColored{62}0.31 & \ValuesColored{62}0.31 & \ValuesColored{62}0.31 & \ValuesColored{62}0.31 & \ValuesColored{62}0.31 & \ValuesColored{62}0.31 \\
    PL    & \ValuesColored{52}0.26 & \ValuesColored{52}0.26 & \ValuesColored{52}0.26 & \ValuesColored{52}0.26 & \ValuesColored{52}0.26 & \ValuesColored{120}0.6 & \ValuesColored{52}0.26 & \ValuesColored{52}0.26 & \ValuesColored{116}0.58 & \ValuesColored{52}0.26 & \ValuesColored{114}0.57 & \ValuesColored{52}0.26 & \ValuesColored{52}0.26 & \ValuesColored{52}0.26 & \ValuesColored{28}0.14 & \ValuesColored{52}0.26 & \ValuesColored{52}0.26 \\
    RO    & \ValuesColored{4}0.02 & \ValuesColored{0}0 & \ValuesColored{0}0 & \ValuesColored{0}0 & \ValuesColored{4}0.02 & \ValuesColored{0}0 & \ValuesColored{0}0 & \ValuesColored{0}0 & \ValuesColored{0}0 & \ValuesColored{4}0.02 & \ValuesColored{0}0 & \ValuesColored{4}0.02 & \ValuesColored{4}0.02 & \ValuesColored{0}0 & \ValuesColored{6}0.03 & \ValuesColored{4}0.02 & \ValuesColored{4}0.02 \\
    SI    & \ValuesColored{142}0.71 & \ValuesColored{112}0.56 & \ValuesColored{146}0.73 & \ValuesColored{146}0.73 & \ValuesColored{100}0.5 & \ValuesColored{146}0.73 & \ValuesColored{146}0.73 & \ValuesColored{146}0.73 & \ValuesColored{146}0.73 & \ValuesColored{144}0.72 & \ValuesColored{146}0.73 & \ValuesColored{142}0.71 & \ValuesColored{144}0.72 & \ValuesColored{146}0.73 & \ValuesColored{146}0.73 & \ValuesColored{144}0.72 & \ValuesColored{142}0.71 \\
    SK    & \ValuesColored{68}0.34 & \ValuesColored{122}0.61 & \ValuesColored{120}0.6 & \ValuesColored{168}0.84 & \ValuesColored{68}0.34 & \ValuesColored{168}0.84 & \ValuesColored{184}0.92 & \ValuesColored{122}0.61 & \ValuesColored{164}0.82 & \ValuesColored{68}0.34 & \ValuesColored{162}0.81 & \ValuesColored{68}0.34 & \ValuesColored{68}0.34 & \ValuesColored{110}0.55 & \ValuesColored{38}0.19 & \ValuesColored{68}0.34 & \ValuesColored{68}0.34 \\
		    \midrule
		AT    & \ValuesColored{152}0.76 & \ValuesColored{132}0.66 & \ValuesColored{168}0.84 & \ValuesColored{164}0.82 & \ValuesColored{152}0.76 & \ValuesColored{164}0.82 & \ValuesColored{146}0.73 & \ValuesColored{106}0.53 & \ValuesColored{164}0.82 & \ValuesColored{158}0.79 & \ValuesColored{166}0.83 & \ValuesColored{152}0.76 & \ValuesColored{60}0.3 & \ValuesColored{162}0.81 & \ValuesColored{168}0.84 & \ValuesColored{158}0.79 & \ValuesColored{152}0.76 \\
    
    BE    & \ValuesColored{66}0.33 & \ValuesColored{24}0.12 & \ValuesColored{118}0.59 & \ValuesColored{88}0.44 & \ValuesColored{66}0.33 & \ValuesColored{144}0.72 & \ValuesColored{66}0.33 & \ValuesColored{52}0.26 & \ValuesColored{32}0.16 & \ValuesColored{66}0.33 & \ValuesColored{36}0.18 & \ValuesColored{66}0.33 & \ValuesColored{66}0.33 & \ValuesColored{84}0.42 & \ValuesColored{200}1 & \ValuesColored{66}0.33 & \ValuesColored{66}0.33 \\
    CH    & \ValuesColored{58}0.29 & \ValuesColored{24}0.12 & \ValuesColored{56}0.28 & \ValuesColored{102}0.51 & \ValuesColored{58}0.29 & \ValuesColored{100}0.5 & \ValuesColored{88}0.44 & \ValuesColored{22}0.11 & \ValuesColored{20}0.1 & \ValuesColored{0}0 & \ValuesColored{54}0.27 & \ValuesColored{58}0.29 & \ValuesColored{40}0.2 & \ValuesColored{68}0.34 & \ValuesColored{86}0.43 & \ValuesColored{0}0 & \ValuesColored{58}0.29 \\
    DE    & \ValuesColored{58}0.29 & \ValuesColored{38}0.19 & \ValuesColored{136}0.68 & \ValuesColored{68}0.34 & \ValuesColored{38}0.19 & \ValuesColored{180}0.9 & \ValuesColored{38}0.19 & \ValuesColored{38}0.19 & \ValuesColored{28}0.14 & \ValuesColored{38}0.19 & \ValuesColored{42}0.21 & \ValuesColored{38}0.19 & \ValuesColored{38}0.19 & \ValuesColored{52}0.26 & \ValuesColored{64}0.32 & \ValuesColored{38}0.19 & \ValuesColored{38}0.19 \\
    DK    & \ValuesColored{56}0.28 & \ValuesColored{108}0.54 & \ValuesColored{34}0.17 & \ValuesColored{32}0.16 & \ValuesColored{40}0.2 & \ValuesColored{54}0.27 & \ValuesColored{20}0.1 & \ValuesColored{20}0.1 & \ValuesColored{56}0.28 & \ValuesColored{0}0 & \ValuesColored{68}0.34 & \ValuesColored{40}0.2 & \ValuesColored{0}0 & \ValuesColored{40}0.2 & \ValuesColored{92}0.46 & \ValuesColored{0}0 & \ValuesColored{40}0.2 \\
    ES    & \ValuesColored{10}0.05 & \ValuesColored{10}0.05 & \ValuesColored{6}0.03 & \ValuesColored{18}0.09 & \ValuesColored{0}0 & \ValuesColored{8}0.04 & \ValuesColored{10}0.05 & \ValuesColored{6}0.03 & \ValuesColored{8}0.04 & \ValuesColored{0}0 & \ValuesColored{0}0 & \ValuesColored{0}0 & \ValuesColored{0}0 & \ValuesColored{8}0.04 & \ValuesColored{0}0 & \ValuesColored{0}0 & \ValuesColored{0}0 \\
		FR    & \ValuesColored{90}0.45 & \ValuesColored{178}0.89 & \ValuesColored{50}0.25 & \ValuesColored{116}0.58 & \ValuesColored{48}0.24 & \ValuesColored{130}0.65 & \ValuesColored{142}0.71 & \ValuesColored{92}0.46 & \ValuesColored{136}0.68 & \ValuesColored{162}0.81 & \ValuesColored{62}0.31 & \ValuesColored{48}0.24 & \ValuesColored{104}0.52 & \ValuesColored{124}0.62 & \ValuesColored{156}0.78 & \ValuesColored{152}0.76 & \ValuesColored{48}0.24 \\
    GB    & \ValuesColored{50}0.25 & \ValuesColored{66}0.33 & \ValuesColored{94}0.47 & \ValuesColored{14}0.07 & \ValuesColored{34}0.17 & \ValuesColored{80}0.4 & \ValuesColored{0}0 & \ValuesColored{60}0.3 & \ValuesColored{62}0.31 & \ValuesColored{0}0 & \ValuesColored{20}0.1 & \ValuesColored{34}0.17 & \ValuesColored{0}0 & \ValuesColored{14}0.07 & \ValuesColored{88}0.44 & \ValuesColored{0}0 & \ValuesColored{34}0.17 \\
    IE    & \ValuesColored{54}0.27 & \ValuesColored{54}0.27 & \ValuesColored{54}0.27 & \ValuesColored{54}0.27 & \ValuesColored{54}0.27 & \ValuesColored{22}0.11 & \ValuesColored{54}0.27 & \ValuesColored{54}0.27 & \ValuesColored{100}0.5 & \ValuesColored{54}0.27 & \ValuesColored{126}0.63 & \ValuesColored{54}0.27 & \ValuesColored{54}0.27 & \ValuesColored{54}0.27 & \ValuesColored{174}0.87 & \ValuesColored{54}0.27 & \ValuesColored{54}0.27 \\
    IT    & \ValuesColored{156}0.78 & \ValuesColored{186}0.93 & \ValuesColored{168}0.84 & \ValuesColored{128}0.64 & \ValuesColored{132}0.66 & \ValuesColored{166}0.83 & \ValuesColored{152}0.76 & \ValuesColored{174}0.87 & \ValuesColored{164}0.82 & \ValuesColored{182}0.91 & \ValuesColored{154}0.77 & \ValuesColored{72}0.36 & \ValuesColored{200}1 & \ValuesColored{162}0.81 & \ValuesColored{146}0.73 & \ValuesColored{176}0.88 & \ValuesColored{28}0.14 \\
    LU    & \ValuesColored{72}0.36 & \ValuesColored{38}0.19 & \ValuesColored{94}0.47 & \ValuesColored{50}0.25 & \ValuesColored{50}0.25 & \ValuesColored{70}0.35 & \ValuesColored{44}0.22 & \ValuesColored{48}0.24 & \ValuesColored{74}0.37 & \ValuesColored{30}0.15 & \ValuesColored{76}0.38 & \ValuesColored{50}0.25 & \ValuesColored{66}0.33 & \ValuesColored{104}0.52 & \ValuesColored{98}0.49 & \ValuesColored{32}0.16 & \ValuesColored{50}0.25 \\
    NL    & \ValuesColored{50}0.25 & \ValuesColored{28}0.14 & \ValuesColored{30}0.15 & \ValuesColored{10}0.05 & \ValuesColored{36}0.18 & \ValuesColored{42}0.21 & \ValuesColored{16}0.08 & \ValuesColored{32}0.16 & \ValuesColored{54}0.27 & \ValuesColored{8}0.04 & \ValuesColored{54}0.27 & \ValuesColored{36}0.18 & \ValuesColored{0}0 & \ValuesColored{28}0.14 & \ValuesColored{76}0.38 & \ValuesColored{100}0.5 & \ValuesColored{36}0.18 \\
    PT    & \ValuesColored{2}0.01 & \ValuesColored{0}0 & \ValuesColored{0}0 & \ValuesColored{4}0.02 & \ValuesColored{0}0 & \ValuesColored{0}0 & \ValuesColored{0}0 & \ValuesColored{0}0 & \ValuesColored{0}0 & \ValuesColored{0}0 & \ValuesColored{0}0 & \ValuesColored{0}0 & \ValuesColored{0}0 & \ValuesColored{0}0 & \ValuesColored{0}0 & \ValuesColored{0}0 & \ValuesColored{0}0 \\
    SE    & \ValuesColored{78}0.39 & \ValuesColored{96}0.48 & \ValuesColored{60}0.3 & \ValuesColored{62}0.31 & \ValuesColored{64}0.32 & \ValuesColored{88}0.44 & \ValuesColored{18}0.09 & \ValuesColored{0}0 & \ValuesColored{90}0.45 & \ValuesColored{64}0.32 & \ValuesColored{106}0.53 & \ValuesColored{64}0.32 & \ValuesColored{64}0.32 & \ValuesColored{74}0.37 & \ValuesColored{128}0.64 & \ValuesColored{64}0.32 & \ValuesColored{64}0.32 \\
    \bottomrule
\end{tabu}%
\end{scriptsize}
\end{table}

\begin{table}[htbp]
\caption{Country abbreviations.}
\label{tab:P2_CountryAbbr}
\centering \small \renewcommand\arraystretch{1.5}
\begin{tabu} to 0.9\linewidth {lXlXlX}
		\multicolumn{2}{l}{\textbf{Eastern \ac{EU}}} 	& \multicolumn{2}{l}{\textbf{Western \ac{EU}}} 	& \multicolumn{2}{l}{\textbf{Non-EU}} \\
		\multicolumn{2}{l}{\textbf{consumers}} 				& \multicolumn{2}{l}{\textbf{consumers}} 				& \multicolumn{2}{l}{\textbf{suppliers}} \\
			BG& \BG{}& AT& \AT{}& AZ& \AZ{}\\ 
			CZ& \CZ{}& BE& \BE{}& DZ$^L$& \DZ{} (LNG)\\ 
			EE& \EE{}& CH& \CH{}& DZ$^N$& \DZ{} (Pipeline)\\
			FI& \FI{}& DE& \DE{}& EG& \EG{}\\
			GR& \GR{}& DK& \DK{}& LY& \LY{}\\ 
			HR& \HR{}& ES& \ES{}& NG& Nigeria\\
			HU& \HU{}& FR& \FR{}& NO$^L$& \NO{} (LNG)\\
			LT& \LT{}& GB& \UK{}& NO$^N$& \NO{} (Pipeline)\\
			LV& \LV{}& IE& \IE{}& OM& \OM{}\\
			PL& \PL{}& IT& \IT{}& PE& \PE{}\\ 
			RO& \RO{}& LU& \LU{}& QA& \QA{}\\
			SI& \SI{}& NL& \TNL{}& RU& \RU\\
			SK& \SK{}& PT& \PT{}& TT& \TT{}\\ 
			& 		 & SE& \SE{}& YE& \YE{}\\  
\end{tabu}
\end{table}

\clearpage
\section{Full set of results} \label{app:P2_ExcelTable}
Appendix\_Results\_full.xlsx contains the full set of results of our simulations. The six sheets of the file display the consumer and producer surpluses, the prices, and the consumption both in absolute and relative terms. Table \ref{tab:P2_RowsAndColumns} describes the columns, which are identical on each sheet of the file. The cell colors highlight the largest positive and negative changes.

{\small \singlespacing \renewcommand\arraystretch{1.5}
\begin{longtabu}  to 0.9\textwidth{lX}
\caption{Legend for Appendix\_Results\_full.xlsx. Columns: indicates the columns of the Excel file. Description: indicates the possible entries in the respective column in italics, and describes the meaning thereof.}
\label{tab:P2_RowsAndColumns} \\
\toprule
\textbf{Column} & \textbf{Description}  \\
\midrule 
\endfirsthead
\multicolumn{2}{c}{\begin{footnotesize}\tablename\ \thetable\ -- \textit{Continued from previous page}\end{footnotesize}} \\
\toprule
\textbf{Column} & \textbf{Description}  \\
\midrule
\endhead
\bottomrule \multicolumn{2}{r}{{\footnotesize\textit{Continued on next page}}} \\
\endfoot
\bottomrule
\endlastfoot
A 		& \textbf{Index} (for sorting purposes) \\ 
\midrule
B 		& \textbf{Simulation year}\\ 
\midrule
C			& \textbf{Simulation $\ID$}\\*
			& \textit{$\ID = 0$:} This line contains the reference simulation.\\
			& \textit{$\ID \in \{1, \ldots, 3\}$:} These lines contain results relative to $\SIM{y,0}$\\
			& \textit{$\ID \in \{4, \ldots, k_y\}$:} These lines contain results relative to $\SIM_{y,3}$\\
\midrule
D 		& \textbf{Simulation type}  \\*
			& \textit{PL:} All production and liquefaction capacities are updated to the current year levels\\
			& \textit{C:}  All values for demand and \wtp{} are updated to the current year levels\\
			& \textit{PLC:} All production and liquefaction capacities, and all values for demand and \wtp{} are updated to the current year level \\
			& \textit{R:} A regasification terminal expansion\\
			& \textit{S:} A storage facility expansion\\
			& \textit{A:} A pipeline expansion\\
			& \textit{RSA:} All regasification, storage, and pipeline capacities are updated to the current year levels \\
			& \textit{RSA \& PLC:} All parameters are updated to the current year levels \\ 
\midrule
E 		& \textbf{Reference value for comparison}\\*
			& $\SIM$ 0: Values given in column D are relative to $\SIM{y,0}$\\
			& $\SIM$ 3: Values given in column D are relative to $\SIM{y,3}$\\ 
\midrule
F 		&\textbf{(Start) location}\\ 
\midrule
G 		& \textbf{End location}\\ 
\midrule
H/J/L & \textbf{Original capacity}. \\*
			&	For storage: injection/extraction/volume capacity\\ 
\midrule
I/K/M & \textbf{Added capacity}. \\*
			&	For storage: injection/extraction/volume capacity\\ 
\midrule
N-AN 	& \textbf{Changes in \cs{} (\textit{dCS}), prices (\textit{dpiC}), and consumption (\textit{dsC})} in the respective countries, see legend of \mbox{Table \ref{tab:P2_CVvalues}} for the country codes. Relative and absolute changes are given on the different sheets of the file.\\ 
\midrule
AO-BS & \textbf{Changes in \ps{} (\textit{dPS})} in the respective countries, see legend of \mbox{Table \ref{tab:P2_CVvalues}} for the country codes. Relative and absolute changes are given on sheets ``Consumer and Producer Surplus'' and ``Consumer and Producer Surplus \%'', respectively.\\ 
\midrule
BT-CA & (Only for sheets ``Consumer and Producer Surplus'' and ``Consumer and Producer Surplus \%'')\\ 
			& \textit{dCS:} Change in \cs{} (in the EU; We only analyze wholesale markets in the EU and therefore cannot give results for \cs{} changes outside the EU\\
			& \textit{dPS EU:} Change in \ps{} within the \ac{EU}\\
			& \textit{dSW EU:} Change in \sw{} within the \ac{EU} (\textit{dSW EU = dCS + dPS EU})\\
			& \textit{dPS:} Change in \ps{} (inside and outside of the EU)\\
			& \textit{dSW:} Change in \sw{} including foreign producers (\textit{dSW = dCS + dPS})\\
			& \textit{dCS abs sum:} Sum of all absolute changes of \cs{}\\
			& \textit{dSW EU abs sum:}	Sum of all absolute changes of \sw{} in the \ac{EU}\\
			& \textit{dSW tot abs sum:} Sum of all absolute changes of \sw{}\\
\end{longtabu}
}
\clearpage
\clearpage
\bibliography{00_Wrapper}
\end{document}